\documentclass[twocolumn,amsmath,amssymb,aps, prx,showkeys,longbibliography,floatfix]{revtex4-1}
\usepackage{graphicx}
\usepackage{bm}
\usepackage{color,soul}
\usepackage{tabularx}
\usepackage{fullpage,subcaption,float,psfrag,textcomp,gensymb,tikz,siunitx}
\usepackage{booktabs,multirow, dcolumn}
\usepackage{xspace}
\usepackage[mathscr]{euscript}
\usepackage{amsmath}
\usepackage{algorithm}
\usepackage{algpseudocode}
\usepackage{paralist}

\makeatletter
\def\BState{\State\hskip-\ALG@thistlm}
\makeatother

\DeclareSymbolFont{rsfs}{U}{rsfs}{m}{n}
\DeclareSymbolFontAlphabet{\mathscrsfs}{rsfs}
\DeclareMathOperator*{\argminA}{arg\,min} 
\DeclareMathOperator*{\argmaxA}{arg\,max} 

\newcommand{\argmax}{\arg\!\max}

\newcommand{\BS}{\mathbf}
\newcommand{\mD}{\mathcal{D}}

\usepackage[bookmarks=true,pdftoolbar=true]{hyperref}
\hypersetup{%
  pdftitle = {Materials Discovery},
  pdfkeywords = {pdf, hyperref, bookmarks},
  pdfauthor = {Raymundo Arroyave}}
 \usepackage[all]{hypcap} 

\def\etal{\mbox{\it et al.\ }}

\graphicspath{{figures/}}

\begin{document}

	\title{Autonomous Efficient Experiment Design for Materials Discovery with Bayesian Model Averaging}

	\author{Anjana Talapatra$^{1}$}
	\email{anjanatalapatra@tamu.edu}
	\author{S. Boluki$^{2}$}
	\author{T. Duong$^{1}$}
	\author{X. Qian$^{2}$}
	\author{E. Dougherty$^{2}$}
	\author{R. Arr\'{o}yave $^{1,3}$}
	\affiliation{$^{1}$ Department of  Materials Science \& Engineering, TAMU, USA, 77843}
	\affiliation{ $^{2}$ Department of Electrical and Computer Engineering, TAMU, USA, 77843}
	\affiliation{ $^{3}$ Department of Mechanical Engineering, TAMU, USA, 77843}
	\date{\today}

	\begin{abstract}
	The accelerated exploration of the materials space in order to identify configurations with optimal properties is an ongoing challenge. Current paradigms are typically centered around the idea of performing this exploration through high-throughput experimentation/computation. Such approaches, however, do not account for---the always present---constraints in resources available. Recently, this problem has been addressed by framing materials discovery as an optimal experiment design. This work augments earlier efforts by putting forward a framework that efficiently explores the materials design space not only accounting for resource constraints but also incorporating the notion of model uncertainty. The resulting approach combines Bayesian Model Averaging within Bayesian Optimization in order to realize a system capable of autonomously and adaptively learning not only the most promising regions in the materials space but also the models that most efficiently guide such exploration. The framework is demonstrated by efficiently exploring the MAX ternary carbide/nitride space through Density Functional Theory~(DFT) calculations.
	\end{abstract}


	\maketitle

\section{Introduction}
\label{sec:Intro}

\subsection{Motivation}
The accelerated exploration of the Materials Design Space (MDS) has been recognized for more than a decade as a key enabler for potentially transformative technological developments~\cite{holdren2011materials,national2008integrated}.  The development of strategies to integrate simulations and experimental data with expert knowledge is a highly active area of research~\cite{agrawal2016perspective,kalidindi2015materials}. Over time, different methods have been deployed within conventional, human-centric, materials development frameworks for exploration of the MDS, including high-throughput (HT) experimentation and computation. 

Traditional HT experimental~\cite{potyrailo2011combinatorial,suram2015generating,green2017fulfilling} and computational~\cite{curtarolo2013high} approaches, while powerful, have important limitations as they (i) employ hardcoded workflows and lack flexibility to iteratively learn and adapt based on the knowledge acquired to assure balanced exploration and exploitation of the MDS (ii) and tend  to be suboptimal in resource allocation as these approaches generally rely on highly parallelized exploration of the MDS, even in regions that are of low value relative to the objective, or performance metric, that is sought after.

Resource limitation cannot be overlooked as it is often the case that once a bottleneck in HT workflows has been eliminated (e.g. synthesis of ever more expansive materials libraries), another one suddenly becomes apparent (e.g. need for high-resolution characterization of materials libraries). Regardless of how many bottlenecks are eliminated, the fact that ultimately a human must make decisions about what to do with the acquired information implies that HT frameworks face hard limits that will be extremely difficult to overcome. On the computational front, there exist significant fundamental and technological challenges to the (multi-scale) simulation of materials~\cite{voorhees2015modeling} that effectively preclude the HT exploration of MDS beyond the use of (sophisticated) methods---such as DFT-based HT simulations~\cite{curtarolo2013high}---operating at one scale, with relatively small numbers of degrees of freedom.

\begin{figure*}[tbp]
\centering
\includegraphics{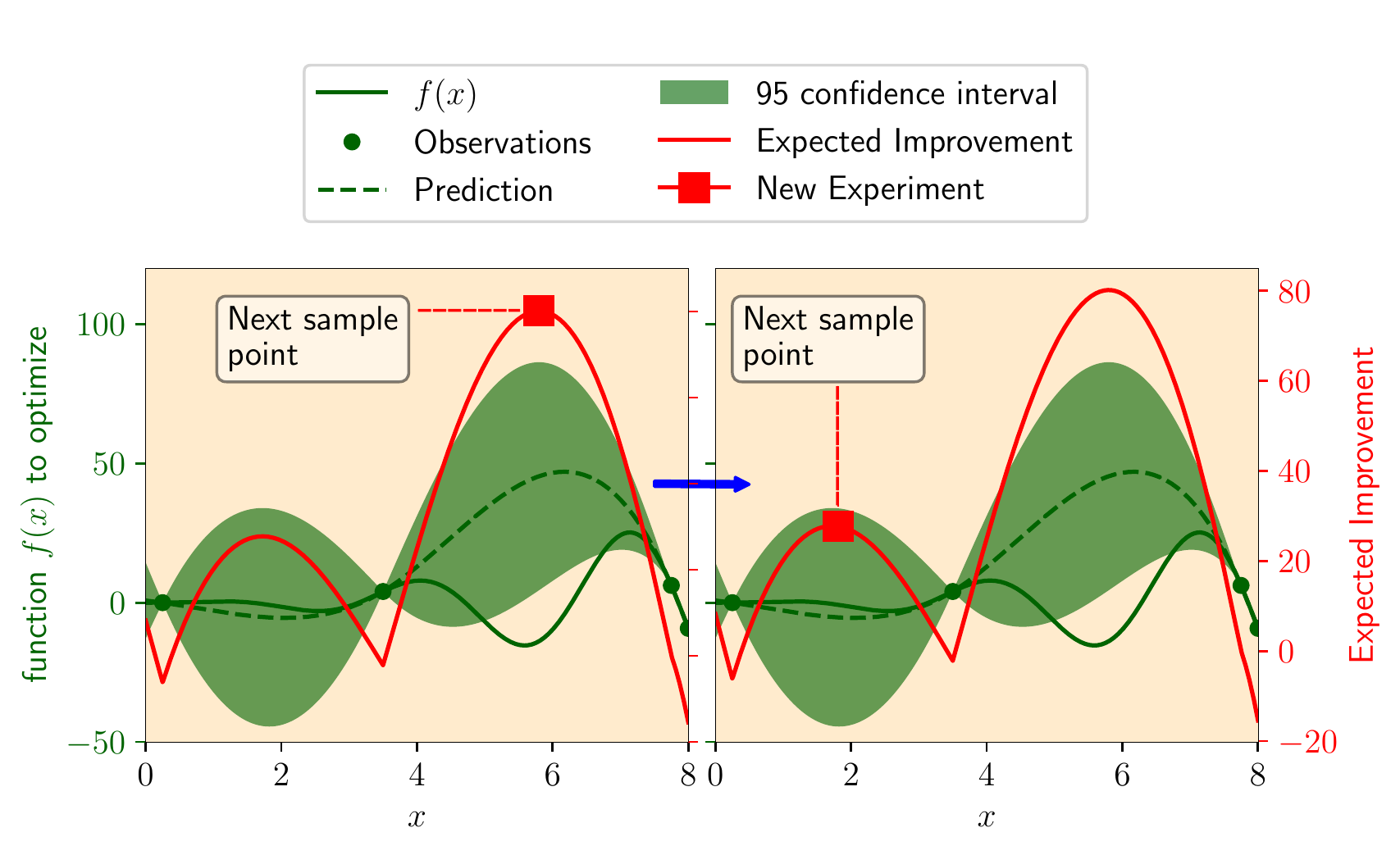}
\caption{Schematic illustration of Bayesian Optimization (BO): from a limited number of observations on a system (blue solid line) a stochastic model (dashed blue line and shaded area) is built. The next observation is determined by accounting for the tradeoff between the exploitation of the current knowledge and the exploration of the unknown regions of the design domain $x$. In this case, Expected Improvement (EI) is the metric used and thus the policy falls within the Efficient Global Optimization (EGO) framework~\cite{jones1998efficient}. }
\label{fig:ego}
\end{figure*}

\subsection{Experiment Design}
\label{sec:addedsubsection}
The goal of any experiment design strategy is to identify an action that
results in a desired property, which is usually optimizing an \textit{%
objective function} of the actions. Without loss of generality, we assume
minimization of the objective function $f\left( \mathbf{x}\right) $: 
\begin{equation}
\mathbf{x}^{\ast }=\arg \min_{\mathbf{x\in }\chi }f\left( \mathbf{x}\right) 
\end{equation}%
\noindent where $\chi $ denotes the action space. In materials discovery,
each action is equivalent to an input or design parameter setup or a
compound, and $\chi $ is the Materials Design Space (MDS). 

The objective function can have a closed form as a parametric function, i.e. 
$f(x,\theta )$, where $\theta $ denotes the parameters. If complete
knowledge of the values of the parameters exist, then no experiments are
needed. In practice, even if a closed form exists, the true values of the
parameters are unknown and they may belong to an uncertainty class $\Theta $%
, governed by a probability measure. Hence, experiments are desired to gain
more knowledge concerning the objective function. It is possible that
the parameters of the objective function are directly parameters of an
underlying system. For example, in \cite{Dehghannasirimocu} the underlying
system is a gene regulatory network and $\theta $ is the set of parameters
that govern the network. In this context, the experiment space can be
different from the action space, e.g. an experiment determines the true
value of a parameter of the underlying system, but an action is a gene
perturbation subsequently determined by a medical criterion dependent on the
value of the parameter. Typically in the context of materials discovery,
each experiment corresponds to applying an action, i.e. setting the input
parameters, and observing its true objective value (or a noisy observation
of it). Whether or not the experiment and action spaces are identical, the
best experiment is determined by optimizing an \textit{acquisition function}.

In materials discovery, $f$ is typically a \emph{blackbox function} without
a known closed form, and the cost of querying such a function (through
expensive experiments/simulations) at arbitrary query point $\mathbf{x}$ in $%
\chi $ is very high. In these cases a surrogate model can be used to
approximate the true objective function. This model can either be parametric
or nonparametric. The so-called Bayesian Optimization (BO) \cite%
{shahriari2016taking} in the literature corresponds to these cases, where
the prior model is sequentially updated after each experiment. Bayesian
parametric and nonparametric models are widely used in other fields such as
bioinformatics \cite{siamak2016,
shahin2017,shahin20171,siamak2017,alireza20181,alireza20182}. When prior
knowledge about the form of the objective function exists and/or many
observations of the objective values at different parts of the input space
are available, one can use a parametric model as a surrogate model. An
example of it for finding the alloy with the least energy dissipation at a
specific temperature can be found at \cite{RoozbehMaterial}, where due to
the availability of the objective values at many initial input points, the
authors have assumed a surrogate parametric function and fixed a subset of
its parameters for the experiment design loop.

If, as is often the case, no prior knowledge of the behavior of the
objective function is available, and limited initial data points are
observed, then one can adopt a nonparametric surrogate model for the
objective function. In either case, there is an inherent feature selection
step, where different potential feature sets might exist. Moreover, there
might be a set of possible parametric families as candidates for the
surrogate model. Even when employing nonparametric surrogate models, several
choices for the kernel functional form might be available. These translate
into different possible surrogate models for the objective function. The
common approach is to select a feature set and a single family of models and
fix this selection throughout the experiment design loop; however, this is
not a reliable approach due to the small initial sample size that is
ubiquitous in materials science. In this paper, we address this problem by
framing experiment design as Bayesian Optimization under Model
Uncertainty (BOMU), and incorporating Bayesian Model Averaging (BMA) within
Bayesian Optimization. Since in the materials discovery context, the
objective function is in most cases a target property of the material;
hereafter the surrogate model for the objective function is referred to as
the \textit{predictive model}.

In the experiments in this paper no prior knowledge about the functional
form of the target properties as functions of the potential features exists,
and Gaussian Process Regression (GPR) \cite{rasmussen2006gaussian} is
employed as the predictive model for each target property. GPR is a flexible
model that imposes only continuity and smoothness prior beliefs and can
asymptotically converge to the true objective function. Moreover, in our
experiments different predictive models correspond to models based on
different potential feature sets. But the approach is by no means limited to
this case and can be applied when different predictive models correspond to
different parametric families or kernel functional forms of nonparametric
models.

A key element in an experiment design strategy is the choice of the acquisition
function. The next selected experiment is the one that maximizes the
acquisition function, which tries to balance the trade-off between the
exploitation of the current belief and the exploration of the unqueried
regions of the input space. The acquisition function is itself dependent on
the modeling of the objective function. Expected Improvement (EI) \cite%
{jones1998efficient} and Knowledge Gradient (KG) \cite{kg2008,correlatedKG}
are among the most commonly used acquisition functions, having been
originally proposed for experiment design under Gaussian belief over the
objective values of input setups and observation noise for an offline
ranking and selection problem. Mean Objective Cost of Uncertainty (MOCU) 
\cite{mocu,Dehghannasirimocu,shahin2018} is another choice for the
acquisition function that is more flexible and quantifies the uncertainty
impacting the operational objectives. For the connection of MOCU with KG and
EGO, the reader can refer to \cite{shahin2018}.

In the following sections, we cover the mathematics of our proposed
algorithm, but the description in words is as follows: 
\begin{compactitem}
\item There is a collection of potential models (e.g. models based on different features sets)
\item The models are averaged, based on the (posterior) model probabilities based on initial data set to form a BMA.
\item Using the expected acquisition function under the BMA, an experiment is chosen that maximizes the expected acquisition.
\item The experiment is run, each model is updated and the (posterior) model probabilities are updated.
\item The expected acquisition under the updated BMA is computed and an experiment is chosen.
\item This iteration is done until some stopping criteria (e.g. while objective not satisfied and budget not exhausted), and the best observation so far is selected as the final suggestion.
\end{compactitem}
In Section \ref{sec:addedappendix} we have provided more details about the generalized MOCU for experiment design and how the approach in this paper compares to that. 
\subsection{Efficient Materials Discovery}
\begin{algorithm}[H]
  \caption{Bayesian Optimization}
  \label{algo:BO}
   \begin{algorithmic}[1]
   \State Initialize $\mathcal{D}_0$
 	 \For{\texttt{n=0,1,... }}
 	    \State Update statistical model $M$
        \State Select new $\mathbf{x}_{\texttt{n}+1}$ by optimizing acquisition function $u$:
        \begin{equation*}
		\mathbf{x}_{\texttt{n}+1}=\underset{\mathbf{x} \in \chi }{\argmaxA}~u\left(\mathbf{x}|\mathcal{D}_\texttt{n},M\right)
	\end{equation*}
      \State Query blackbox function $f$ to obtain $y_{n+1}$
	  
      \State Augment data $\mathcal{D}_{\texttt{n}+1}=\left\{\mathcal{D}_\texttt{n},\left(\mathbf{x}_{\texttt{n}+1},y_{\texttt{n}+1}\right)\right\}$
      \If{stopping criteria reached}
      \State {$\textbf{break}$}
      \EndIf

            \EndFor
   \end{algorithmic}
\end{algorithm}
Resource constraints call for the \emph{efficient} evaluation of materials configurations in order to identify regions in the MDS with the \emph{optimal response}. Bayesian Optimization (BO)~\cite{shahriari2016taking,pelikan1999boa} provides a sequential model-based approach to solve the problem: first, a prior belief is prescribed over the objective function and then the model ($M$) is sequentially refined via Bayesian posterior updating. The domain $\chi$ is sampled for a query point $\mathbf{x}_{\texttt{n}+1}$ such that an acquisition function $u(\mathbf{x}|\mathcal{D}_\texttt{n},M)$---constructed from a model of the observed data  $\mathcal{D}_\texttt{n}$---is maximized--see Algorithm~\ref{algo:BO} and Fig.~\ref{fig:ego}. The stopping criteria can be reaching the desired properties or exhausting the experimental budget.

Having mapped the exploration of the MDS to an expensive \emph{blackbox function}, several groups have already demonstrated the power of Bayesian Optimization in the context of \emph{accelerated} materials discovery. Early on, Fujimura \etal~\cite{fujimura2013accelerated} combined DFT and experimental data to construct a model to predict the ionic conductivity of Li-super ionic conductors via Support Vector Regression (SVR)~\cite{basak2007support}. The predicted conductivity $\sigma_{Li}$ from the SVR model was then used as the \emph{acquisition} function to further explore the Li-super ionic conductor space. Seko \etal ~\cite{seko2014machine} used feature sets derived from DFT calculations and experimentally measured melting points $T_m$ to fit stochastic models based on SVR or Gaussian Process Regression (GPR)~\cite{rasmussen2006gaussian} to discover unary and binary crystals with the highest melting point. In that case, the acquisition function used in the sequential exploration of the melting point space $\chi_{T_m}$ was the \emph{probability} of improving upon the best value recorded before acquisition $n+1$. These early results introduced the notion of sequential exploration but did not consider the larger implications of framing the materials discovery as the optimization of an expensive \emph{blackbox function}.

Balachandran \etal~\cite{balachandran2016adaptive} prescribed the need to balance the need to exploit our current knowledge of the MDS $\chi$ with the need to explore it. The balance between exploitation and exploration was realized by invoking a proper acquisition function. Balachandran \etal proposed using \emph{Expected Improvement}~\cite{jones1998efficient}, $EI$, in the predicted objective function $y$ by the model $P(y|\BS{x},\mD)$ over the unexplored regions of $\chi$, given the observed data $\mathcal{D}$. $EI$ can in turn be calculated for unexplored query points $\mathbf{x}$  by the model trained. They demonstrated their design protocol by attempting to predict the MAX phases (ternary layered carbides/nitrides~\cite{barsoum2013max}) with maximal/minimal polycrystalline bulk/shear moduli as predicted via DFT calculations. Having demonstrated the power of Bayesian Optimization in materials discovery, the same group~\cite{xue2016accelerated} notably employed the same approach to discover, \emph{via experiments}, NiTi-based Shape Memory Alloys (SMAs) with record-low hysteresis through a minimal experimental effort.

The principled nature of a BO-based materials discovery protocol is amenable to develop full-loop platforms, particularly when attempting to carry out simulation-driven materials development. Indeed, Ju \etal~\cite{ju2017designing} recently proposed a framework whereby atomistic transport calculations were combined with a BO framework to identify aperiodic nano-structures with optimal transport properties by examining only an extremely small fraction of the possible configurations. On the experimental front, Nikolaev \etal~\cite{nikolaev2016autonomy} recently demonstrated a fully autonomous closed-loop iterative materials experimentation platform. They demonstrated the system by optimizing the synthesis conditions for carbon nanotubes. In their case, the approach focused on a \emph{greedy} exploitation of the synthesis space by using the predicted rate of growth as the acquisition function---i.e. no exploitation-exploration tradeoff~\cite{xue2016accelerated,balachandran2016adaptive} was used.

\subsection{Contributions of this Paper}

While existing computational and experimental deployments of optimal materials discovery constitute significant advances, there are still significant challenges that remain to be addressed. For example, most BO-based approaches rely on a feature selection step~\cite{balachandran2016adaptive,lookman2016statistical,jalem2018bayesian} that necessarily requires a considerable number of feature-property sets to be effective~\cite{broderick2015informatics}. In other cases, the strength of the approach depends on building sufficient prior knowledge (from informative predictive models~\cite{seko2014machine}) in order for \emph{greedy} approaches to be practical. 

Unfortunately, more often than not, the amount of relevant data available before embarking on a materials discovery problem is small. In such situations the nature (and dimensionality) of the design space---$\chi$ in the BO formalism---is not known \emph{a priori}. Moreover, it is not even clear which features are best connected to the target performance metric. Finally, the inability of existing approaches to `\emph{both build and exploit their internal models, with minimal human hand-engineering}'~\cite{botvinick2017building} precludes the implementation of truly autonomous materials discovery systems, even in simulation-driven approaches.

In this work, we propose a framework that, simultaneously (i) accounts for the need to adaptively build increasingly effective models for the accelerated discovery of materials while (ii) accounting for the uncertainty in the models themselves. The framework is then demonstrated by efficiently exploring the MAX ternary carbide space through DFT calculations. Incorporating BMA within Bayesian Optimization produces a system capable of autonomously and adaptively learning not only the most promising regions in the materials space but also the models that most efficiently guide such exploration. The framework is also capable of defining optimal experimental sequences in cases where multiple objectives must be met---we note that recent works have begun to address the issue of multi-objective Bayesian Optimization~\cite{gopakumar2018multi} in the context of materials discovery. Our approach, however, is different in that the multi-objective optimization is carried out simultaneously with feature selection.

\section{Bayesian Optimization under Model Uncertainty}
Small sample sizes are ubiquitous in materials science. Experiments---and simulations---are often resource-intensive and this imposes significant constraints on any attempt to explore/exploit the MDS. Moreover, in the absence of sufficient information, there are, \emph{a priori}, multiple features that are potentially predictive of the material performance metric of interest. In all the well-known experiment design methods in the literature, one must select the model (the set of predictive features and/or the parametric form or the kernel functional form of the model) before starting the experiment design loop. 

Unfortunately, due to small sample size and large number of potential predictive models, the model selection step may not result in the true best predictive model for efficient Bayesian Optimization~\cite{DoughertySima,DoughertyUlisses}. It has been shown that small sample sizes pose a great challenge in model selection due to inherent risk of imprecision and overfitting \cite{DoughertySima,DoughertyUlisses}, and no feature selection method performs well in all scenarios when sample sizes are small \cite{Dougherty2009FeatureSelection}. Thus, by selecting a single model as the predictive model based on small observed sample data, one ignores the model uncertainty \cite{Raftery7page1996BMA}. 

\subsection{Building Robust Predictive Models through Bayesian Model Averaging}
\label{sub:BMA}

One possible approach to circumvent this problem is to weight all the possible models by their corresponding probability of being the true model, and use all of these in the experiment design step so that model uncertainty can be taken care of for Bayesian Optimization. In other words, the derived predictive model is a marginalized aggregation of all the potential predictive models, weighted by the prior probability and likelihood of the observed data for that model, resulting in the Bayesian Model Averaging (BMA) method~\cite{RafteryTutorial1999BMA,wasserman2000BMA}.

Here, we discuss the multi-output case from which the single output can be readily deduced. Let $y^j$ represent the $j^{\text{th}}$ output of interest, and $\BS{x}$ the corresponding vector of features or materials design parameters, and the observed data be denoted by  $\mD=\{\BS{X},\BS{Y}\}$, where $\BS{Y}=[Y^1,...,Y^q]$ is a matrix having the collection of the observed $j^{\text{th}}$ output as its $j^{\text{th}}$ column, i.e. $Y^j=[y^j_1,...,y^j_n]^T$, where $n$ is the number of observed data points, and $\BS{X}$ represent the matrix of the collection of the corresponding observed features. Here, to simplify the notation we have dropped the subscript denoting the experiment iteration step for $\mD$, but note that $\mD=\mD_\texttt{n}$ at any $n$th step. The predictive probabilistic model for $\BS{y}$ for a new feature vector $\BS{x}$ after observing $\mD$ is
\begin{equation}
P(\BS{y}|\BS{x},\mD) = \sum_{i=1}^{L}P(M_i|\mD)P(\BS{y}|\BS{x},\mD,M_i),
\end{equation}
where $P(\BS{y}|\BS{x},\mD,M_i)$ represents each potential probabilistic predictive model, and 
\begin{align}
&P(M_i|\mD) = \allowbreak \frac{P(\mD|M_i)P(M_i)}{\sum_{j=1}^{L}P(\mD|M_j)P(M_j)},\label{eq:posterior-model-prob}\\
&P(\mD|M_i)=\int P(\mD|\BS{\theta}_i,M_i)P(\BS{\theta}_i|M_i)d\BS{\theta}_i,\label{eq:marginal-data-likelihood}
\end{align}
\noindent are the (posterior) probability of each model being the true predictive model, and the marginal probability of the observed data under model $M_i$, respectively. $L$ is the total number of models under consideration, and $M_i$ and $\BS{\theta}_i$ represents the $i^{\text{th}}$ model and the vector of $i^{\text{th}}$ model parameters, respectively. 

If we further assume independence among outputs and let $\mD_j$ denote $\{\BS{X},Y^j\}$, we have $P(\BS{y}|\BS{x},\mD,M_i)=\prod_{j=1}^{q}P(y^j|\BS{x},\mD_j,M_i)$ and
\begin{equation}
\begin{split}
P(\mD|M_i) & = \prod_{j=1}^{q} P(\mD_j|M_i) \\
                   &=\prod_{j=1}^{q}\int P(\mD_j|\BS{\theta}^j_i,M_i)P(\BS{\theta}^j_i|M_i)d\BS{\theta}^j_i.
\end{split}
\end{equation}

When each potential probabilistic predictive model $M_i$ is a Gaussian Process Regression (GPR) model \cite{Rasmussen:2005:GPM:1162254}, $\BS{\theta}^j_i$ are the parameters of the covariance function. In fact, each GPR model $M_i$ is defined by a mean (basis) function ($m^j_i(\cdot)$) and a covariance function ($K^j_i(\cdot,\cdot;\BS{\theta}_i^j)$). In this setup, $P(y^j|\BS{x},\mD,M_i)$ is a Gaussian distribution, i.e. $P(y^j|\BS{x},\mD,M_i)=\mathcal{N}(\mu^j_i(\BS{x}),\sigma_i^{2,j}(\BS{x}))$, where the predicted mean and variance of the $j^{\text{th}}$ objective function are \cite{Rasmussen:2005:GPM:1162254}:
\begin{align}\label{eq:gpr-predict-oneObj}
&\mu^j_i(\BS{x}) = m^j_i(\BS{x})+ \notag \\& \quad \quad K^j_i(\BS{x}, \BS{X};\BS{\theta}^j_i)K^j_i(\BS{X},\BS{X};\BS{\theta}^j_i)^{-1}(Y^j-m^j_i(\BS{X})), \notag\\
&\sigma_i^{2,j}(\BS{x}) =K^j_i(\BS{x}, \BS{x};\BS{\theta}^j_i)-\notag \\ & \quad \quad K^j_i(\BS{x}, \BS{X};\BS{\theta}^j_i)K^j_i(\BS{X}, \BS{X};\BS{\theta}^j_i)^{-1}K^j_i(\BS{X}, \BS{x};\BS{\theta}^j_i).
\end{align}

In practice, when using type II maximum likelihood (ML-II) estimation, the covariance function parameters of each model are estimated by maximizing the marginal log-likelihood of the observed data under that model, i.e. an estimate $\hat{\BS{\theta}}^j_i$ is calculated by maximizing
\begin{widetext}
\begin{equation}\label{eq:marginal-likelihood-oneObj}
\text{log}P(D_j|\BS{\theta}^j_i,M_i) = -\frac{1}{2}(Y^j-m^j_i(\BS{X}))^TK^j_i(\BS{X},\BS{X};\BS{\theta}^j_i)^{-1}(Y^j-m^j_i(\BS{X})) - \frac{1}{2}|K^j_i(\BS{X},\BS{X};\BS{\theta}^j_i)| - \frac{n}{2}\text{log}2\pi, 
\end{equation}
\end{widetext}
where $|\cdot|$ denotes matrix determinant. A quasi-Newton method with multiple random starts can be employed to find the maximum of \eqref{eq:marginal-likelihood-oneObj}. This estimate $\hat{\BS{\theta}}^j_i$ is then used in \eqref{eq:gpr-predict-oneObj} for prediction purposes under the model assumptions.

For a GPR, $P(\mD_j|\BS{\theta}^j_i,M_i)$ is a multivariate Gaussian probability density function, and $P(\mD_j|M_i)=\int P(\mD_j|\BS{\theta}_i^j,M_i)P(\BS{\theta}_i^j|M_i)d\BS{\theta}_i^j$, the marginal probability of the observed data corresponding to $j^{\text{th}}$ output under model $M_i$ in \eqref{eq:marginal-data-likelihood}, can be approximated by
either first-order expansion of the exponent, or second-order expansion of the exponent known as Laplace approximation method~\cite{Rasmussen:2005:GPM:1162254}. In the first-order approximation, since $\hat{\BS{\theta}}_i^j$ is a stationary point of \eqref{eq:marginal-likelihood-oneObj}, $P(\mD_j|M_i)$ can be approximated by $P(\mD_j|\hat{\BS{\theta}}_i^j,M_i)$. In the second-order approximation, $P(\mD_j|M_i)\approx P(\mD_j|\hat{\BS{\theta}}_i^j,M_i)\int \text{exp}\big(-\frac{1}{2}(\BS{\theta}_i^j - \hat{\BS{\theta}}_i^j)^T (-H(\hat{\BS{\theta}}_i^j)) (\BS{\theta}_i^j - \hat{\BS{\theta}}_i^j)\big)d\BS{\theta}_i^j$, where $H(\hat{\BS{\theta}}_i^j)$ is the Hessian matrix of $\text{log}P(\mD_j|\BS{\theta}_i^j,M_i)$ calculated at $\hat{\BS{\theta}}_i^j$. When all the models are assumed to have the same probability \emph{a priori}, the posterior model probabilities in \eqref{eq:posterior-model-prob}, i.e. $P(M_i|\mD),i=1,...,L,$ are only dependent on the marginal probability of the observed data under each model in \eqref{eq:marginal-data-likelihood}, i.e. $P(\mD|M_i),i=1,...,L$.

\subsection{Experiment Design by Bayesian Optimization}

Bayesian Experiment Design (BED) has the potential to guide efficient search for desired materials by directing sequential search of ``optimal'' query points to approach the optimal solution~\cite{shahriari2016taking}. Here, we employ the Expected Improvement (EI) \cite{jones1998efficient} for single objective problems, and an extension of EI to guide the search to approach the Pareto front for multi-objective problems, namely the Expected Hyper-Volume Improvement (EHVI) \cite{emmerich2011hypervolume}. Both EI and EHVI can balance exploration and exploitation up to some extent in guiding the search for optimal solutions. 

A \emph{major innovation} in our BED approach is that instead of assuming knowledge of the best predictive model in advance and updating this given predictive model based on the limited number of initial observed data and iterating the experiment design loop based on the updated model---an approach that is taken in the literature---we consider the model uncertainty by including a class of potential predictive models for the task under study. By BMA, the experiment design step is performed based on the weighted average of these potential models. After performing the selected experiment, the new observed data from the experiment is used to update the (posterior) probability of all these potential predictive models. We can see that by taking this approach, as more experiments are done, the true predictive model is selected with a higher probability alongside accelerating the discovery of the material with the desired properties.We note that the proposed BMA also works in cases in which the feature sets are known or fixed but in which different model forms of the GPR---i.e. using different kernels---could potentially have different degrees of fidelity with regards to the available data.

For multi-objective problems, the EHVI under model averaging is

\begin{equation}\label{eq:ehvi-averaging}
\begin{split}
&EI_{\mathcal{H}}(\BS{x}|\mD) = \int I_{\mathcal{H}}(\BS{y}|\BS{x},\mD)P(\BS{y}|\BS{x},\mD)d\BS{y} = \\ & \int I_{\mathcal{H}}(\BS{y}|\BS{x},\mD) \sum_{i=1}^{L}P(M_i|\mD)P(\BS{y}|\BS{x},\mD,M_i)d\BS{y} = \\ & \sum_{i=1}^{L}P(M_i|\mD)EI_{\mathcal{H}}(\BS{x}|\mD,M_i),
\end{split}
\end{equation}
where $I_{\mathcal{H}}(\BS{y}|\BS{x},\mD)$ denotes the hyper-volume improvement achieved by observing the outputs at $\BS{x}$, and $EI_{\mathcal{H}}(\BS{x}|\mD,M_i)$ is the ordinary EHVI under model $M_i$. If the outputs are assumed to be independent $EI_{\mathcal{H}}(\BS{x}|\mD,M_i)$ further simplifies to $\int I_{\mathcal{H}}(\BS{y}|\BS{x},\mD)\prod_{j=1}^{q}P(y^j|\BS{x},\mD,M_i)d\BS{y}$. The optimal experiment to be performed next is $\BS{x}^* = 
 \underset{\mathbf{x} \in \mathcal{X} }{\argmax}~EI_{\mathcal{H}}(\BS{x}|\mD)$, which is the one that maximizes the weighted average EHVI considering all the potential predictive models, again by the iteratively updated (posterior) model probabilities. The hyper-volume improvement $I_{\mathcal{H}}(\BS{y}|\BS{x},\mD)$ is the increase in the hyper-volume of the dominated (objective) space achieved by adding the outputs at $\BS{x}$ to the observed data, i.e. $I_{\mathcal{H}}(\BS{y}|\BS{x},\mD)=\mathcal{H}(\BS{Y}\cup \BS{y})-\mathcal{H}(\BS{Y})$. Without loss of generality, if we assume the goal is minimization of all the outputs, the hyper-volume dominated by a set of points $\BS{A}$ is defined as the volume of the dominated subspace by the points in $A$, i.e. $\mathcal{H}(\BS{A}) =\mathrm{Volume} \left( \{\boldsymbol{s}\in \mathbb{R}^q|\boldsymbol{s}\prec \boldsymbol{r}\, \text{and}\, \exists\, \BS{a} \in \BS{A}: \, \BS{a} \prec \boldsymbol{s}\}\right)$, where the domination rule is such that $\BS{a}\prec \BS{b}$ if and only if $a^j\leq b^j$ for all $j=1,...,q$, and for at least one $j$, $a^j < b^j$. $\BS{r}$ is called a reference or anchor point and is a point dominated by all the possible output values (the whole output space).
 
For the special case of employing EI-based BED~\cite{jones1998efficient}, the EI after observing data $\mD$ can be computed under model averaging by:  

\begin{equation}\label{eq:ei-averaging}
\begin{split}
EI(\BS{x}|\mD) & = \int I(y|\BS{x},\mD) \sum_{i=1}^{L}P(M_i|\mD)P(y|\BS{x},\mD,M_i)dy\\ 
                        &= \sum_{i=1}^{L}P(M_i|\mD)\int I(y|\BS{x},\mD)P(y|\BS{x},\mD,M_i)dy\\
                        & =\sum_{i=1}^{L}P(M_i|\mD)EI(\BS{x}|\mD,M_i),
\end{split}
\end{equation}
where $I(y|\BS{x},\mD)$ denotes the improvement achieved by observing the output of experiment $\BS{x}$, $E$ represents expectation, and $EI(\BS{x}|\mD,M_i)$ is the EI under model $M_i$. In this approach, the optimal experiment to be performed next is $\BS{x}^* = \underset{\BS{x} \in \chi }{\argmax}~EI(\BS{x}|\mD)$. We can see that the optimal experiment is the one that maximizes the weighted average EI considering all the potential predictive models based on the iteratively updated (posterior) model probabilities given the observed data. In the equations above, the improvement achieved by observing the output of experiment $\BS{x}$ is $I(y|\BS{x},\mD)=(y^*-y)_+$ when minimization is the goal, and $I(y|\BS{x},\mD)=(y-y^*)_+$ when maximization is the goal, where $(a)_+=a$ if $a>0$ and is zero otherwise, and $y^*$ denotes the best (lowest/highest for minimization/maximization problems) output observed so far, i.e. the best output in $\mD$. 

\begin{figure*}[tbp]
\centering
\includegraphics[width=0.75\textwidth]{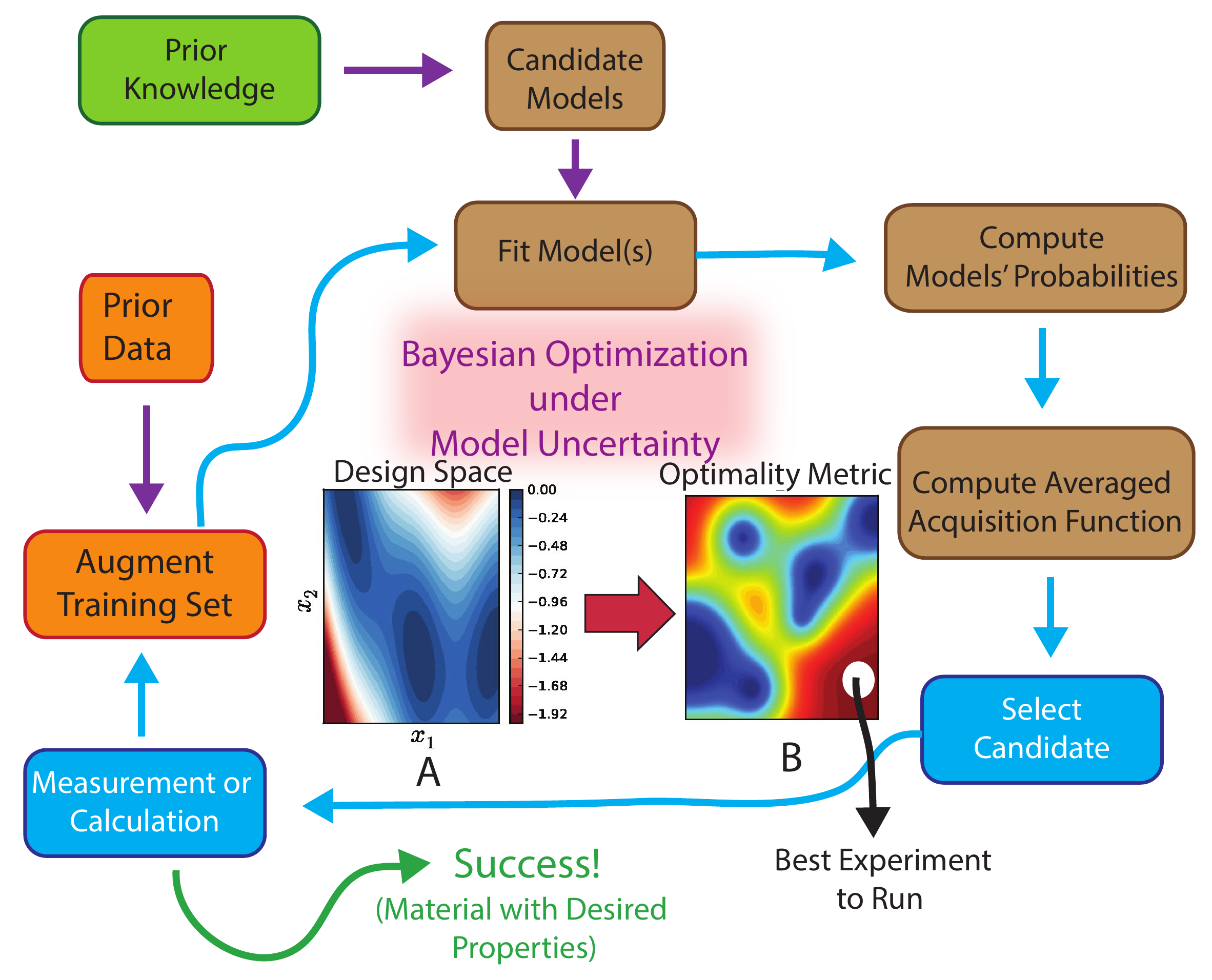}
\caption{Schematic of the proposed framework for an autonomous, efficient materials discovery system as a realization of Bayesian Optimization under Model Uncertainty (BOMU). Initial data and a set of candidate models are used to construct a stochastic representation of an experiment/simulation. Each model is evaluated in a Bayesian sense and its probability is determined. Using the model probabilities, an effective acquisition function is computed, which is then used to select the next point in the materials design space that needs to be queried. The process is continued iteratively until target is reached or budget is exhausted.}
\label{fig:bomu}
\end{figure*}

For the GPR model assumptions taken by the experiments in this paper, we have chosen the constant mean function (i.e. $m_i(\BS{x})=c_i$ for single output and $m^j_i(\BS{x})=c^j_i$ for multiple output cases) and the (Gaussian) Radial Basis Function~(RBF) kernel, a popular choice, for the covariance function: 
\begin{equation}
K_i^j(\BS{x}, \BS{x'};\BS{\theta}_i^j) = \theta^j_{i,1} \text{exp}\left[ -\frac{1}{2} \frac{\|\BS{x}-\BS{x'}\|^2}{\theta^j_{i,2}}\right].
\end{equation}

The focus of the experiments in this paper is on showing the power of experiment design considering model uncertainty by BMA in guiding the search towards the optimal compound (with corresponding features design parameters) when the best predictive model is not known in advance, a usual case in practical applications, while also identifying the best predictive model as more data from experiments become available. The algorithm for our proposed Bayesian Optimization under Model Uncertainty (BOMU) framework is shown in Algorithm~\ref{algo:BOMU} and the overall framework for autonomous materials discovery is shown in Fig.~\ref{fig:bomu}. In Algorithm~\ref{algo:BOMU}, for the single-objective case, $u(\BS{x}|\mD_\texttt{n},M_i)$ and $u(\BS{x}|\mD_\texttt{n})$ correspond to $EI(\BS{x}|\mD_\texttt{n},M_i)$ and $EI(\BS{x}|\mD_\texttt{n})$, and for the multi-objective case correspond to $EI_{\mathcal{H}}(\BS{x}|\mD_\texttt{n})$ and $EI_{\mathcal{H}}(\BS{x}|\mD_\texttt{n},M_i)$, respectively.

In this paper we consider predictive models based on different potential feature sets. The details are provided in the following section.

\begin{algorithm}[H]
  \caption{Bayesian Optimization under Model Uncertainty}
  \label{algo:BOMU}
   \begin{algorithmic}[1]
   \State Initialize $\mathcal{D}_0$
 	 \For{\texttt{n=0,1,... }}
        \State Update statistical model(s), $M_i$
        \State Compute acquisition function $u$ with model averaging:
        \begin{equation*}
        u\left(\mathbf{x}|\mathcal{D}_\texttt{n}\right)=\sum_{i=1}^{L}P(M_i|\mD_\texttt{n})u(\BS{x}|\mD_\texttt{n},M_i)
        \end{equation*}	 
        
        \State Select new $\mathbf{x}_{\texttt{n}+1}$ by optimizing acquisition function $u$:
        \begin{equation*}
		\mathbf{x}_{\texttt{n}+1}=\underset{\mathbf{x} \in \chi }{\argmaxA}~u\left(\mathbf{x}|\mathcal{D}_\texttt{n}\right)
	\end{equation*}
      \State Query blackbox function $f$ to obtain $y_{n+1}$
      \State Augment data $\mathcal{D}_{\texttt{n}+1}=\left\{\mathcal{D}_\texttt{n},\left(\mathbf{x}_{\texttt{n}+1},y_{\texttt{n}+1}\right)\right\}$
      \If{stopping criteria reached}
      \State {$\textbf{break}$}
      \EndIf

            \EndFor
   \end{algorithmic}
\end{algorithm}

\section{Deployment of BOMU: Optimal Discovery of the MAX Phase Space}

\begin{figure*}[htb]
\centering
\includegraphics[width=0.99\textwidth]{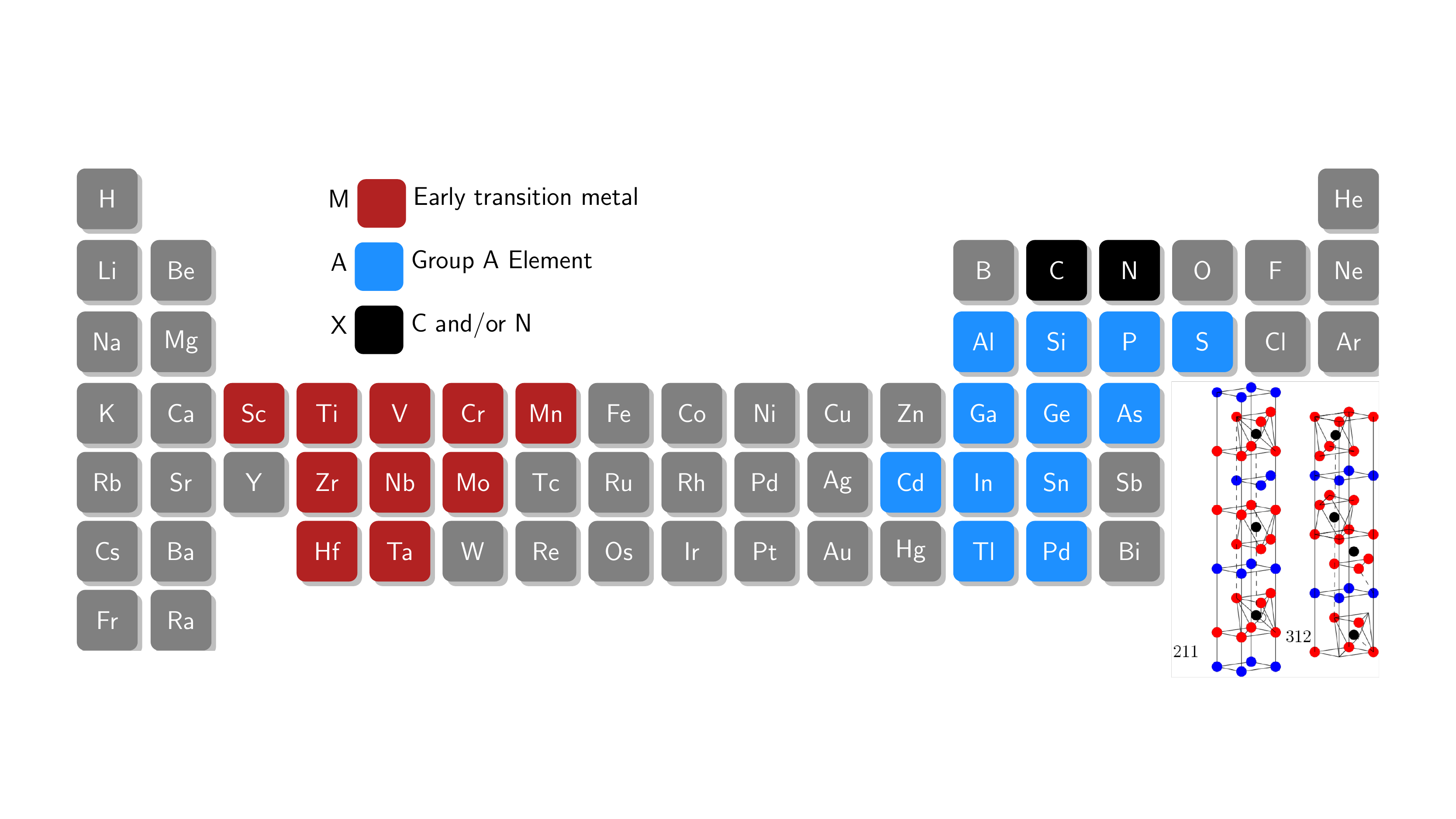}
\caption{MAX Phases: ternary (and higher order) layered carbides and nitrides with properties intermediate to those of metals and ceramics~\cite{barsoum2013max}.}
\label{fig:max}
\end{figure*}

M$_{n+1}$AX$_n$ phases---$M$ corresponds to a transition metal, $A$ corresponds to group IV and VA elements, and $X$ corresponds to carbon or nitrogen---have a property range within those of ceramics and metals due to the coexistence of both metallic and metallic/covalent bonds within their layered structures~\cite{barsoum2013max, radovic2013max, barsoum2000, barsoum2004,barsoum2011elastic,sun2011progress}. The bonds between M-A layers tend to be much weaker than those between M-X layers, making them easily deformable while retaining much of the chemical (and thermodynamic) stability of transition MX carbides. While only a very small fraction of the pure ternary MAX phase composition palette has been synthesized to date~\cite{barsoum2010max}, there is a considerable opportunity to uncover promising chemistries with optimal property sets once different stacking sequences and deviations from stoichiometries in the M, A, and X sites are considered~\cite{arroyave2016does,talapatra2016high}.

\subsection{Design Problem: Optimal Mechanical Properties in the MAX Phase Space}
Because of their rich chemistry and the wide range of values of their properties~\cite{aryal2014genomic}, MAX phases constitute an adequate material system to test simulation-driven---specifically DFT calculations---materials discovery frameworks. Aryal \etal~\cite{aryal2014genomic}, for example, carried out an exhaustive investigation of the structural, electronic and stability properties of 792 MAX phases with the M$_{n+1}$AX$_n$ and $n=$1-4. Balachandran \etal~\cite{balachandran2016adaptive} used the MAX phases with M$_{2}$AX stoichiometry to deploy and test different Bayesian Optimization schemes. In this work, we use the same system to test the proposed framework. 

 The MDS for this work is composed of conventional MAX phases with  $M_{2}AX$ and $M_{3}AX_2$ stoichiometries. Here M $\in$ $\{Sc, Ti, V, Cr, Zr, Nb, Mo, Hf, Ti \}$; A $\in$ $\{ Al, Si, P, S, Ga, Ge, As, Cd, In, Sn, Tl, Pd \}$; and X $\in$ $\{C, N\}$. This results in 216 $M_{2}AX$ and  216 $M_{3}AX_2$ phases. Since we are testing a materials discovery framework, we found it convenient to determine the ground truth of the system beforehand and the mechanical properties of these systems were thus determined before deploying the BOMU framework ---our framework has been incorporated into a high-throughput workflow automation tool using the scikit-learn \cite{pedregosa2011scikit} toolbox. The implementation is publicly available at \href{https://gitlab.com/tammal/matpredict}{https://gitlab.com/tammal/matpredict}. Of the possible MAX phases with the chemistries described above, 29 were found to be elastically unstable and were discarded. The design space thus consists of 403 MAX phases.
 
The problem was formulated with the goal of identifying the material/materials with i) the maximum bulk modulus $K$; ii) the minimum shear modulus $G$; and iii) the maximum bulk modulus and minimum shear modulus.
The cases of i) the maximum bulk modulus $K$; ii) the minimum shear modulus $G$ are designed as single-objective optimization problems. The third problem which seeks to identify the materials with the maximum bulk modulus and minimum shear modulus (iii) is designed as a multi-objective problem.

\subsection{Prior Knowledge}
In this framework, it is assumed that some prior knowledge is available before starting the materials discovery task. This prior knowledge could be as simple as a set of features that are likely to have effects on the materials properties of interest. Here we describe the features that were selected to constitute the MDS:

Each candidate MAX phase $c_i$ of the MDS is determined uniquely by a set of features $F_j$, which have been distilled from the existing literature and the authors' domain knowledge. In the MDS, the $M$ \& $A$ chemical elements comprising the MAX phases sweep along the rows and columns of the periodic table and it is reasonable to expect that features which intrinsically reflect periodic trends would characterize the properties of the MAX phases.  

In this work, a total of fifteen features were considered: empirical constants $C, m $ which relate the elements comprising the material to it's bulk modulus \cite{makino2000estimation}; valence electron concentration $C_v$; electron to atom ratio $\frac{e}{a}$; lattice parameters $a$ and $c$; atomic number $Z$; interatomic distance $I_{dist}$; the groups according to the periodic table of the M, A \& X elements $Col_M, Col_A, Col_X $ respectively; the order $O$ of MAX phase (whether of order 1 corresponding to $M_2AX$ or order 2 corresponding to $M_3AX_2$); the atomic packing factor $APF$; average atomic radius $rad$; and the volume/atom $vol$. In relevant cases ($C,m,Cv, \frac{e}{a},Z,I_{dist}, APF, C_v$), these features are composition-weighted averages calculated from the elemental values and are assumed to propagate as per the Hume-Rothery rules. 

The $ C, m $ parameters are related to the bonding character. These are composition-weighted values of the empirical constants reported by Makino \emph{et al.}\cite{makino2000estimation}, who proposed that the bulk modulus $\mathbf K$ of elemental substances can be determined by the relation $K = Cr_{ps}^{-m}$; where $r_{ps}$ is the effective pseudopotential radius. The valence electron concentration $C_v$ is another feature related to the bonding character and is a known marker of the stability of a phase \cite{karthikeyan2015role,Guo_VEC}. The $\frac{e}{a}$ ratio, which is the average number of itinerant electrons per atom, plays a significant role in the bonding of a solid and is closely related to the valence electron concentration $C_v$ \cite{mizutani2016determination}.  

The lattice parameters $c,a$ for all the domain elements were calculated by DFT by allowing the structures to completely relax. The $c$ lattice parameter is highly correlated to the order of the MAX phase (whether $M_2AX$ or $M_3AX_2$). The lattice parameters implicitly account for the effect of volume and atomic radius on the elastic properties. Additionally, the $c/a$ ratio characterizes the MAX phases, they being hexagonal close packed (hcp) materials. The relationship between the elastic properties and the $c/a$ ratio for hcp materials has also been extensively studied \cite{pronk2003large,tromans2011elastic}. Here we note that since the determination of the equilibrium structural parameters is approximately an order of magnitude less costly than the full calculation of the elastic constant tensor and thus it is a reasonable proposition to use these DFT-derived quantities to assist in the prediction/discovery of properties that are more costly to acquire.

The atomic number $Z$, which denotes the number of electrons is the foremost factor that determines the chemical bonding behavior of a material and defines its chemical properties. The weighted interatomic distance $I_{dist}$ was calculated from the elemental values, which were sourced from the CRC Handbook of Chemistry and Physics \cite{lide1998crc}. The atomic packing factor (APF) plays an important role in the determination of elastic properties. For example, face centered cubic (fcc) structures tend to be ductile, while hcp structures are brittle. Finally, the  structural parameters: average atomic radius $rad$ and the volume/atom $vol$ were determined from the DFT-determined lattice parameters.

\subsection{Determining Candidate Models}

As discussed, the determination of features comprising the MDS was based off of prior literature and domain knowledge. \textit{A priori}, it is not known which of these features significantly influence the target properties in the materials discovery task.
In the search for new materials with desired properties, such situations are often encountered, where there is a lack of fundamental knowledge relating the intrinsic nature of the material and the desired property. 
The BOMU approach invoked in this work accounts for uncertainty in the models $M_i$ available to fit the blackbox predictive model to observed data. In our design problem, different models $M_i$ correspond to different subsets $\mathcal{F}_S$ out of the entire feature set $\mathcal{F}$, $\mathcal{F}_S~\subseteq~\mathcal{F}$.  

While one could question the need to define candidate feature subsets $\mathcal{F}_S$ when the entire feature set $\mathcal{F}$ is available, it is important to note that exploring the entire feature set is problematic due to important limitations~\cite{kandasamy2015high}. First, non-parametric regression is challenging in high-dimensional space, with lower bounds of nearest-neighbour distance between samples depending exponentially on the dimension of the problem~\cite{gyorfi2006distribution}. This exponential complexity affects the convergence rate of BO approaches~\cite{bull2011convergence}. Second, the computational effort in maximizing the acquisition function also increases in an exponential manner with the number of features. 

The general problem of Bayesian Optimization in the presence of many potential models (feature sets) is still an outstanding challenge~\cite{wang2016bayesian} and different approaches have been proposed, including the partitioning of the domain in disjoint subdomains~\cite{kandasamy2015high} or the use of random embedding~\cite{wang2016bayesian}. No approach so far provides the means for the BO framework itself to `learn' the optimal model and select the subspace most effectively to reach the target property(ies). Our proposed approach, as will be shown below, \textbf{addresses these issues} and thus constitutes a novel approach to effectively reduce the complexity of the BO problem under model uncertainty.

\subsection{Selecting Feature Sets}

Feature selection is an essential component of model construction and learning and is a research area in itself. Application of rigorous feature selection methods can lead to better models with a good understanding of the underlying characteristics of the data. Using the right features reduces 
the complexity of the model and reduces overfitting. Choosing the right subset of features also improves the accuracy of a model. For the purposes of this work,  we elected to see how far one can get by  choosing to rely on simpler methods. To reduce the feature space dimensionality of the model, we grouped the features into 6 sets containing 4 features each, as shown in Table~\ref{tab:features}. Of the 15 features considered, only 13 were used, with $rad$ and $vol$ being discarded.

\begin{table}[]
\centering
\caption{Feature Sets Considered in this Design}
\label{tab:features}
\begin{tabular}{ll}
\hline\hline
$F_1$            & $[ C, m, C_v, c]$                         \\
$F_2$            & $[m, Z, I_{dist}, \frac{e}{a}]$          \\
$F_3$            & $[ \frac{e}{a}, a, c, C_v ]$             \\
$F_4$            & $[ C, m, C_v, Col_A]$                    \\
$F_5$           & $[Col_M, Col_A, Col_X, O]$            \\
$F_6$            & $[a, c, APF, I_{dist}]$                  \\ \hline
\end{tabular}
\end{table}

These sets were created \emph{adhoc}, using a combination of physical insights and an effort to make sets containing features which reflect the effect of electronic structure and chemical bonding character. For example, since $C$ and $m$ are derived from Makino's empirical model \cite{makino2000estimation}, they were grouped together in sets $F_1$ and $F_4$. In set $F_2$, $m$ was used standalone, since the empirical relationship $K = Cr_{ps}^{-m}$ indicates that $m$ is more significant than $C$, which only introduces the effect of a constant. In set $F_5$, only the compositional element markers ($Col_M, Col_A, Col_X$) along with the order $O$ of MAX phase were used, to simulate a feature set which has only the most basic compositional and structural description.

\subsection{Materials Discovery/Design Protocol}

GPR models based on six feature subsets in Table~\ref{tab:features}, were adopted in our BMA experiment design. For each of the targets (maximizing K, minimizing G, as well as maximizing-K/minimizing G) we carried out the sequential experiment design by maximizing the EI or EHVI based on predictive models using single feature sets or BMA using all the feature sets accounting for their probability through first-order (BMA$_1$) and second-order (BMA$_2$) Laplace approximation.

The optimization scheme was run for initial data sets ( i.e known data points) of size $N=2, 5, 10, 15 ,20 $. The `training set' thus ranges from $\approx 1/200$ to $1/20$ of the MDS. 1500 instances of each initial set $N$ were used to ensure a stable average response. The budget for the optimal design was set at $\approx 20 \%$ of the MDS, i.e 80 materials or calculations. In each iteration, two calculations were done. The selection for the compound(s) to query is based on the optimal policy used: EI or EHVI. Thus the candidates with the maximum and second maximum EI/EHVI are selected for update. This means that for example, for the maximization of the bulk modulus problem for N=2, we initially know the bulk modulus of 2 materials (N=2) and can calculate the bulk modulus of 78 more materials to stay within the budgeted 80 calculations. Since we are calculating the bulk modulus of 2 materials at a time, this means a total of 78/2 = 39 iterations for this case. All the input features were normalized, before being fed to the optimization module. 

\subsection{DFT Calculation Parameters}

The total energy calculations were performed within the DFT~\cite{kohn1965self} framework, as implemented in the \emph{Vienna ab initio} simulation package (VASP)~\cite{vasp,vasp2}. The generalized gradient approximation (GGA)~\cite{GGA} is used in the form
of the parameterization proposed by Perdew, Burke, and Ernzerhof (PBE)~\cite{perdew1996generalized}. Brillouin zone integrations were performed using a Monkhorst-Pack mesh~\cite{monkpack} with at least 5000 k points per reciprocal atom. Full relaxations were realized by using the Methfessel-Paxton smearing method~\cite{smearing} of order one and a final self-consistent static calculation with the tetrahedron smearing method with Bl{\"o}chl corrections~\cite{tetramethod}. A cutoff energy of 533 eV was set for all of the calculations and the spin polarizations were taken into account. 

To estimate the lattice parameters, the structures were allowed to fully relax to their ground states. The relaxations were carried out in six stages: first stage by allowing change in volume ( corresponding to the VASP ISIF =7 tag), second stage by additionally allowing the relaxation of cell shape (corresponding to the VASP ISIF =6 tag), third stage by also allowing relaxation of ions (corresponding to the VASP ISIF =3 tag), fourth stage by allowing only the ions to relax (corresponding to the VASP ISIF =2 tag), fifth stage by allowing full relaxation (VASP ISIF =3 tag) and a final self-consistent static calculation run. All relaxations were carried out until changes in total energy between relaxation steps were within $1$ x $10^{-6}$ eV. 

The elastic constants were calculated using the stress-strain approach \cite{le2002symmetry,singh2016effect} where a set of strains ($\epsilon_1; \epsilon_2;\epsilon_3;\epsilon_4;\epsilon_5; \epsilon_6$) were imposed on a crystal, determined using DFT methods as described in \cite{duongelastic}. From the $n$ set of strains and the resulting stresses, elastic constants were calculated based on Hooke's law. For these calculations, the ionic positions were relaxed while leaving the lattice shape and volume invariant. These calculations were followed by a static calculation using order-one Methfessel-Paxton smearing method and an auxiliary FFT grid to ensure maximum accuracy in the calculation of interatomic forces. Convergence criteria ensured that calculated elements of elastic constant tensor changed within a few GPa when varying the magnitude of the lattice strain from 1\% to 3\%. From these elastic constants, various elastic properties have been calculated using the Voigt and Reuss approximations and Voigt-Reuss-Hill averaging \cite{hill1952elastic}. The properties under consideration are: the bulk modulus (K) and the shear Modulus (G).

\section{Results}
\label{sec:results}
As mentioned earlier, we employ the EI and EHVI acquisition functions in the experiment design loop for single and multi-objective problems, respectively. Hereafter, a single model is a Gaussian Process Regression (GPR) model based on a single feature set. Also, $F_1$, $F_2$, $F_3$, $F_4$, $F_5$, and $F_6$ denote the 6 different feature sets considered in our analysis, the GPR models based on those feature sets, and experiment design assuming the underlying model based on those feature sets, interchangeably. In the following ``convergence'' for each model (feature set) refers to the calculation number in the experiment design iterations based on that model (feature set) when the true optimal design parameters are identified in (nearly) all simulations with 1500 initial data sets with different size $N$ for each setup.

\subsection{Single objective optimization}
\subsubsection{Maximization of bulk modulus (K)}

\begin{figure}
\centering
\includegraphics{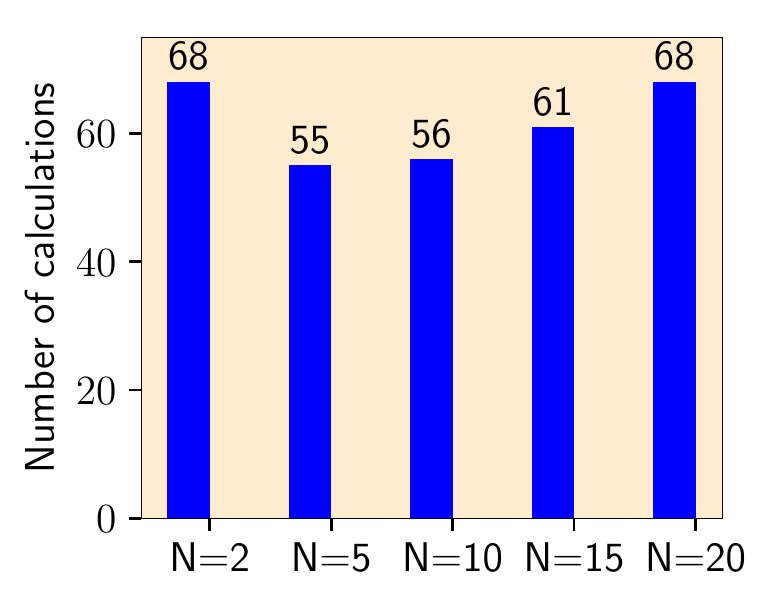}
\caption{ Average number of calculations required to find maximum bulk modulus for different numbers of initial data instances $N = 2,5,10,15,20$ using feature set $F_2$}
\label{fig:F_2_bar_plot}
\end{figure}
\begin{figure*}[h!]
        \parbox{.975\textwidth}{
            \begin{subfigure}{.475\linewidth}
                \includegraphics{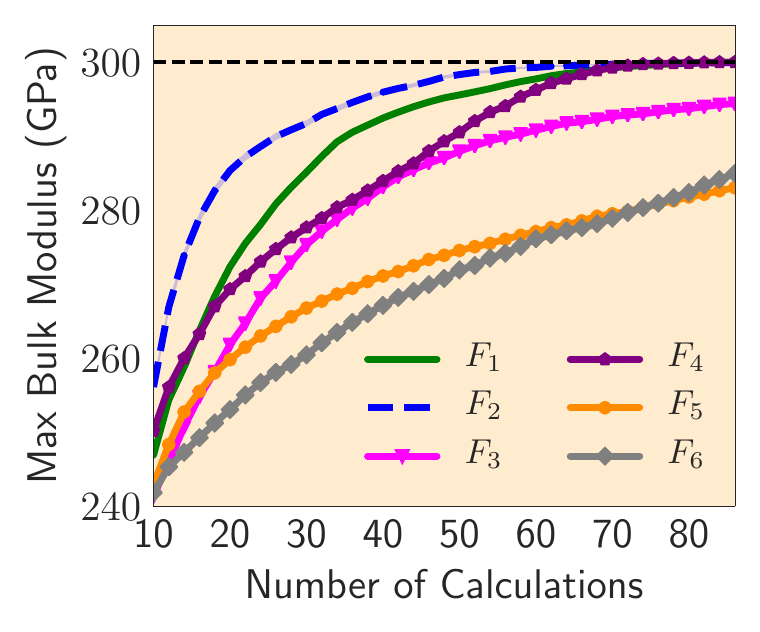}
                \caption{}
                \label{fig:Max_bulk_modulus_single_models}
        \end{subfigure}
        	\begin{subfigure}{.475\linewidth}
                \includegraphics{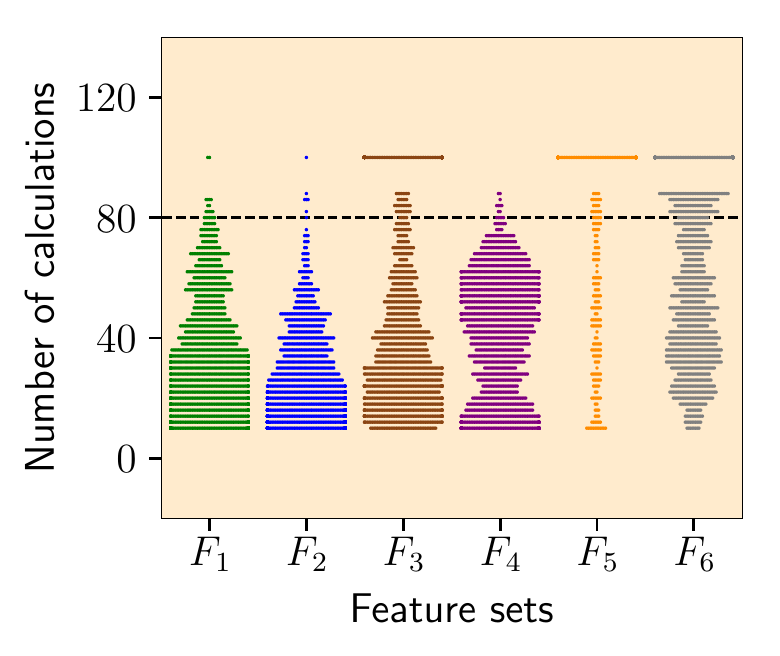}
                \caption{}
                \label{fig:K_max_all_violinplots}
        \end{subfigure}
        	\begin{subfigure}{.475\linewidth}
                \includegraphics{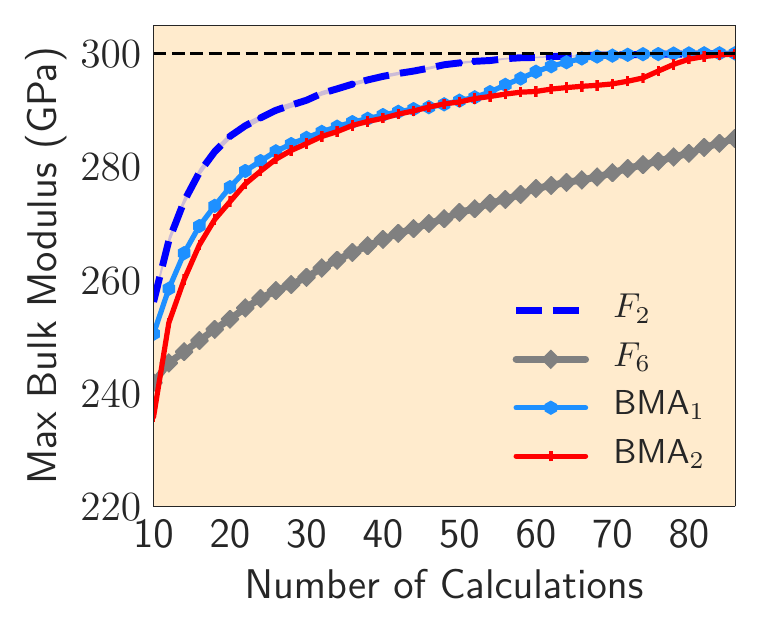}
                \caption{}
                \label{fig:Max_bulk_modulus_BMA}
        \end{subfigure}
        	\begin{subfigure}{.475\linewidth}
                \includegraphics{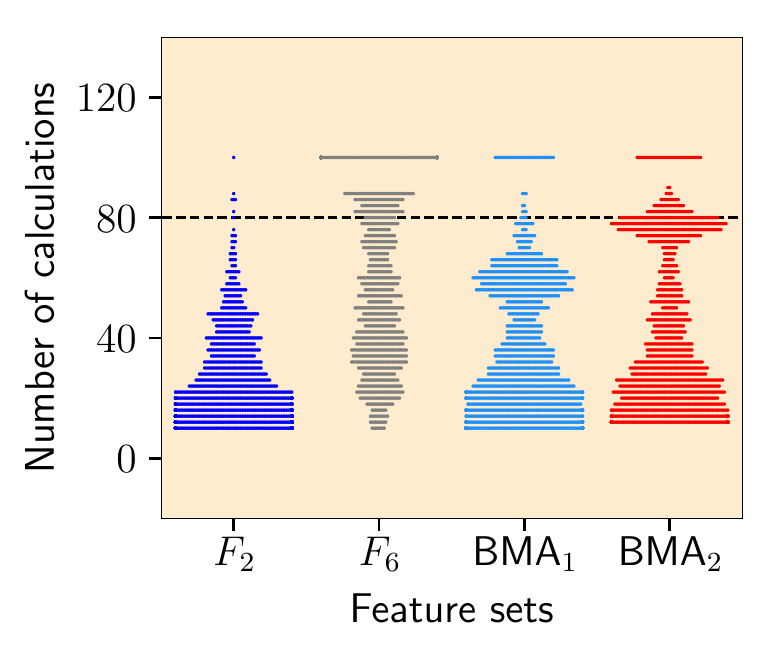}
                \caption{}
                \label{fig:Max_bulk_modulus_BMA_violinplot}
        \end{subfigure}
        }
        \caption{ Representative results for single objective optimization -- maximization of bulk modulus for N=10: a) Average maximum bulk modulus discovered using all described feature sets,  b) swarm plots indicating the distribution of the number of calculations required for convergence using all described feature sets, c) average maximum bulk modulus discovered using the best feature set $F_2$, worst feature set $F_6$, BMA$_1$ and BMA$_2$, and d) swarm plots indicating the distribution of the number of calculations required for convergence using best feature set $F_2$, worst feature set $F_6$, BMA$_1$ and BMA$_2$.}
        \label{fig:Max_bulk_modulus_single_BMA}
    \end{figure*}

 As mentioned earlier, calculations were carried out for different number of initial data instances $N =2,5,10,15,20$. The performance trends for all three problems across different values of $N$ are consistent.
 The technique is found to not significantly depend on quantity of initial data. Fig. \ref{fig:F_2_bar_plot} shows the average number of calculations required to find maximum bulk modulus for $N = 2,5,10,15,20$ with $F_2$. Even when we start with very few initial data instances at $N=2$, the Bayesian experiment design (BED) procedure converges at least as fast as $N=20$. Using $N=5$ however, leads to faster convergence than starting with $N=10,15,20$. 
 This shows that it is often more effective to start with a small initial data set. This is advantageous, since in real-world problems, scarcity of data is a common limitation. Consequently, for the sake of brevity, we present results using the representative case of $N=$10 only. Results for $N=2,5,15,20$ may be found in the Supplementary material \cite{supplementary}. 
{}
For the first test problem to find the MAX phase with the maximum bulk modulus, the maximum values found in the experiment design iterations based on each model (feature set)  averaged over all initial data set instances for $N=$10 are shown in Fig. \ref{fig:Max_bulk_modulus_single_models}. The dotted line in the figure indicates the maximum bulk modulus = 300 GPa that can be found in the MDS. $F_2$ is found to be the best performing feature set on average, converging fastest to the maximum bulk modulus. In other words, using the predicted values as well as uncertainty estimation from the GPR model with $F_2$ in the experiment design loop guides us toward the optimal solution of the problem faster than the other models. $F_6$ and $F_5$ on the other hand, are uniformly the worst performing feature sets on average, converging the slowest. 

Fig. \ref{fig:K_max_all_violinplots} shows the swarm plots indicating the number of calculations required to discover the maximum bulk modulus in the MDS using experiment design based on single models for the 1500 initial data instances with  $N=$10. The width of the swarm plot at every vertical axis value indicates the proportion of instances where the optimal design parameters were found at that number of calculations. Bottom heavy, wide bars, with the width decreasing with the number of steps is desirable, since that would indicate that larger number of instances needed fewer number of steps to converge. The dotted line indicates the budget allotted, which was 80 calculations. Instances that did not converge within the budget were allotted a value of 100. Thus the width of the plots at vertical value of 100, corresponds to the proportion of instances which did not discover the maximum bulk modulus  in the MDS within the budget. From this figure, it is seen that for $F_1, F_2$ and $F_4 $ in almost 100 \% of instances the maximum bulk modulus was identified within the budget, while $F_5$ is the poorest feature set and the maximum was identified in very few instances. 

Fig. \ref{fig:Max_bulk_modulus_BMA} shows the comparison of the average performance of both the first-order and second-order BMA over all initial data set instances with the best performing model ($F_2$) and worst performing model ($F_6$). It can be seen that both the first-order and second-order BMA performance in identifying the maximum bulk modulus is consistently close to the best model ($F_2$). First-order BMA performs as well as if not better than $F_2$. Fig. \ref{fig:Max_bulk_modulus_BMA_violinplot} shows the corresponding swarm plots indicating the number of calculations required to discover the maximum bulk modulus in the MDS for the 1500 instances of initial data set for $N=$10 using first- and second-order BMA, respectively. It can be seen that for a very high percentage of cases the maximum bulk modulus can be found within the designated budget.

\begin{figure*}
        \parbox{.975\textwidth}{
            \begin{subfigure}{.475\linewidth}
                \includegraphics[width=\textwidth]{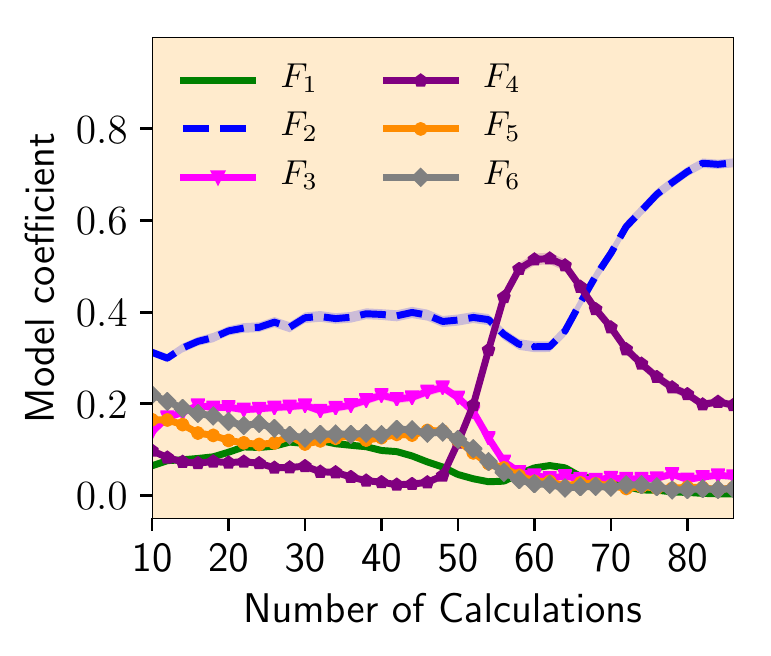}
                \caption{BMA$_1$}
                \label{fig:Max_bulk_modulus_coefficients_BMA1}
        \end{subfigure}
        \begin{subfigure}{.475\linewidth}
                \includegraphics[width=\textwidth]{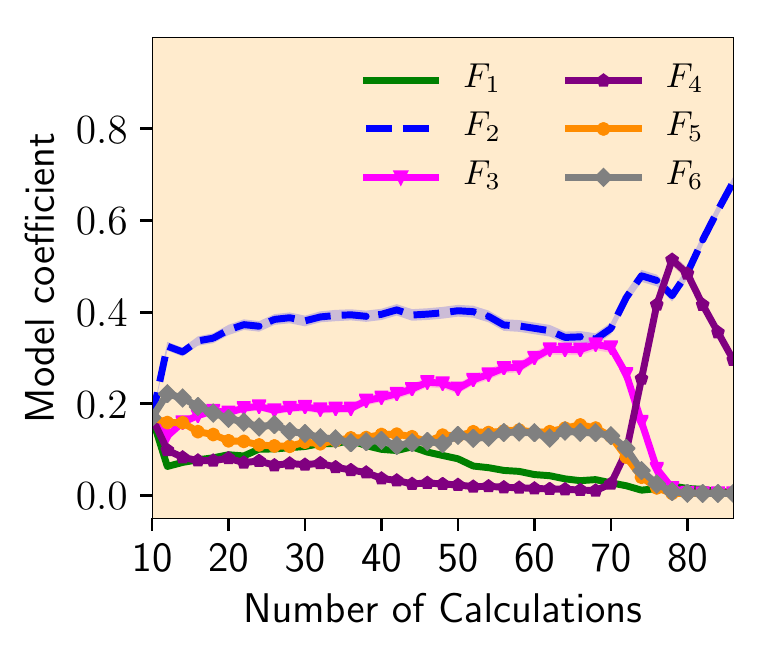}
                \caption{BMA$_2$}
                \label{fig:Max_bulk_modulus_coefficients_BMA2}
        \end{subfigure}
        \caption{Average model probabilities for maximizing bulk modulus using a)BMA$_1$ and b) BMA$_2$}
        \label{fig:Max_bulk_modulus_coefficients_BMA}
        }
\end{figure*} 

\begin{figure}
\centering
\includegraphics[width=0.99\columnwidth]{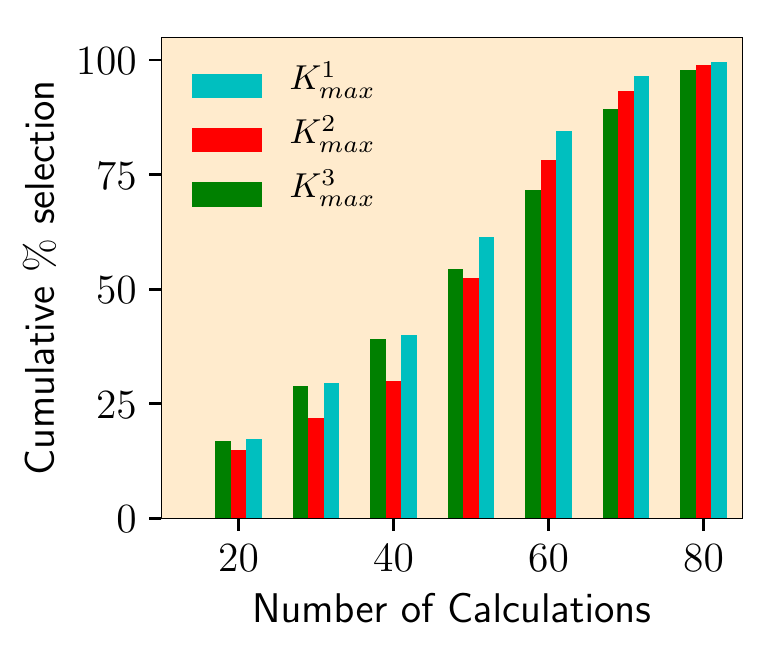}
\caption{ Percentage of BED selected materials with the maximum ( $K_{max}^{1}$), second maximum ( $K_{max}^{2}$), third maximum ( $K_{max}^{3}$) bulk modulus with the increasing number of calculations for BMA$_1$. }
\label{fig:bar_plot}
\end{figure}

In Figs. \ref{fig:Max_bulk_modulus_coefficients_BMA1} and~\ref{fig:Max_bulk_modulus_coefficients_BMA2}, the average model coefficients (posterior model probabilities) of the GPR models based on different feature sets over all instances of initial data set are shown with the increasing number of calculations for BMA$_1$ and BMA$_2$, respectively. It can be seen that these model coefficients from BMA may guide automatic selection of the best feature set $F_2$. For BMA$_1$ and BMA$_2$, the average probability of $F_2$ is (almost) always higher than the other models. Earlier, in Figs. \ref{fig:Max_bulk_modulus_single_models} and~\ref{fig:K_max_all_violinplots}, $F_4$ also appears to be a good model and converges at par with $F_2$ around the 75 calculations. Reflecting this, as the number of available experiments/calculations increases (55 for BMA$_1$ and 75 for BMA$_2$), the model probability of $F_4$ briefly overtakes that of $F2$ as indicated in Fig. \ref{fig:Max_bulk_modulus_coefficients_BMA}.  As more data become available, BMA again considers $F_2$ as the best model based on the updated model coefficients during the experiment design procedure. Note that such a feature set selection based on BMA is directly determined by the performance of achieving desired operational objectives for experiment design. 
The actual candidate materials selected during each progressive BED iteration with BMA$_1$ were analyzed over the 1500 instances, among which the cumulative  percentage of choosing candidates with the maximum ($K_{max}^{1}$), second maximum ($K_{max}^{2}$), third maximum ($K_{max}^{3}$) bulk modulus is indicated in Fig.~\ref{fig:bar_plot}. It is seen that as the BED loop proceeds and the surrogate model improves, the materials with the maximum bulk modulus (top 3 for illustration) are selected more consistently. Specifically, beyond approximately 40 calculations, there is a steep increase in the selection of $K_{max}^{i}$ as a candidate, corresponding to the steep increase in the probability of model $F_4$ and $F_2$ as illustrated in Fig. \ref{fig:Max_bulk_modulus_coefficients_BMA1}.

\subsubsection{Maximization of bulk modulus: Non-informative features}
\label{sec:non_informative_features}
\begin{figure*}
\centering
\begin{subfigure}{.5\textwidth}
  \centering
  \includegraphics[scale=0.8]{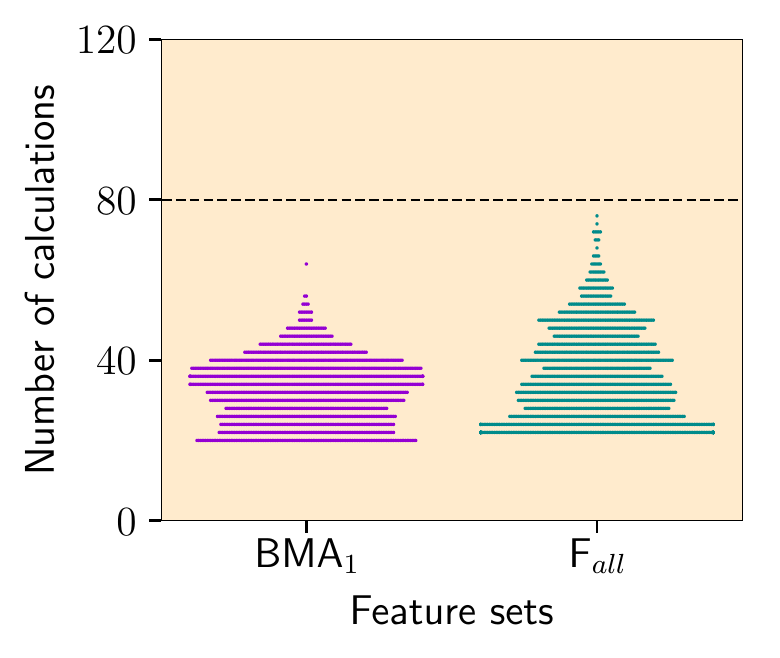}
  \caption{}
  \label{fig:29_coeffs}
\end{subfigure}%
\begin{subfigure}{.5\textwidth}
  \centering
  \includegraphics[scale=0.8]{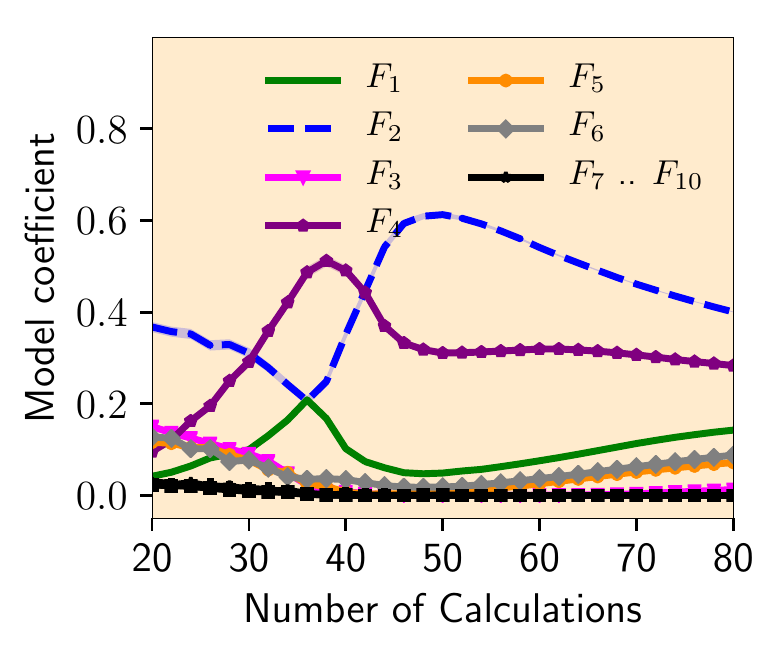}
  \caption{}
  \label{fig:29_swarm}
\end{subfigure}
\caption{Representative results for single objective optimization -- minimization of shear modulus for N=20  for the case of 29 features: a) average model probabilities for maximizing bulk modulus using BMA$_1$ and F$_all$ b) swarm plots indicating the distribution of the number of calculations required for convergence using  BMA$_1$ and F$_{all}$.}
\label{fig:29_features}
\end{figure*}

 To showcase the utility of our BMA approach, we simulate a high-dimensional case by adding 16 non-informative random features, which we compose into subsets F$_7$, F$_8$, F$_9$, and F$_{10}$ of four features each. We carry out two types of calculation using the larger set of 29 (13+16) features. First, we use the $BMA_1$ approach to find material with maximum K using F$_1$,...F$_{10}$; and we use the regular EGO-GP framework to find the material with maximum K using all 29 features. The results for the same are plotted in Figure \ref{fig:29_features}. Firstly, we see in Figure \ref{fig:29_features}a,  that in this case (an actual high dimensional case with a number of non-informative random features), the BMA approach outperforms using all features together. Additionally, tracking the model probabilities as in Figure \ref{fig:29_features}b, shows us that the BMA approach effectively picks up the F$_2$ set as the best feature set, rejects the random feature sets F$_7$, ...F$_{10}$ (average model probabilities are negligible) and performs better than using F$_2$ standalone (in Figure \ref{fig:Max_bulk_modulus_BMA_violinplot}).

\subsubsection{Minimization of shear modulus (G)}

\begin{figure*}
        \parbox{.975\textwidth}{
            \begin{subfigure}{.475\linewidth}
                \includegraphics[width=\textwidth]{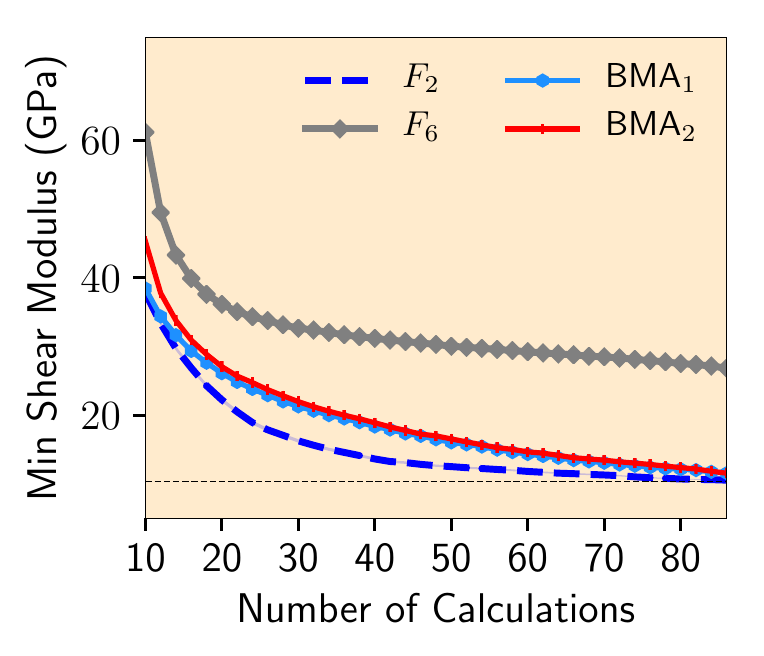}
                \caption{}
                \label{fig:N_10_G_min_BMA}
        \end{subfigure}
        \begin{subfigure}{.475\linewidth}
                \includegraphics[width=\textwidth]{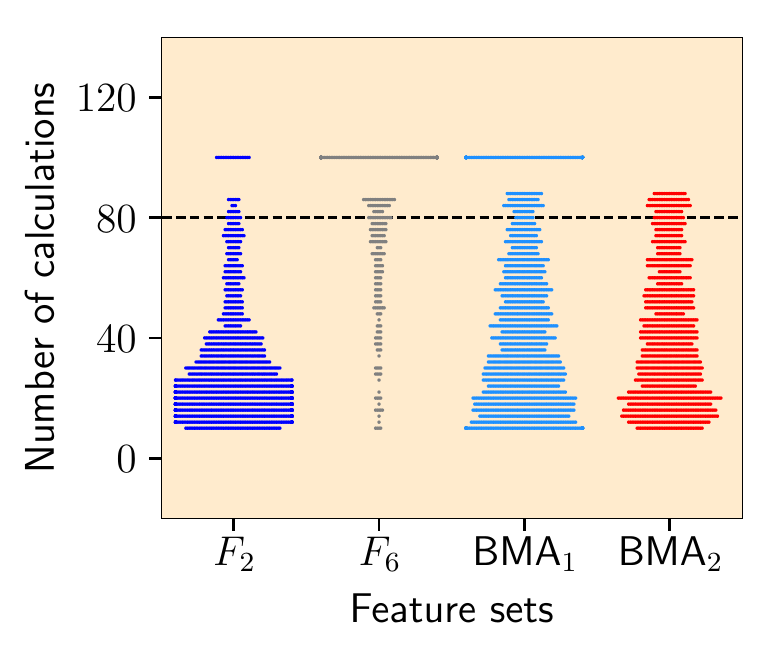}
                \caption{}
                \label{fig:N_10_G_min_BMA_violinplot}
        \end{subfigure}
        }
        \caption{Representative results for single objective optimization -- minimization of shear modulus for N=10: a) average minimum shear modulus discovered using the best feature set $F_2$, worst feature set $F_6$, BMA$_1$ and BMA$_2$, and d) swarm plots indicating the distribution of the number of calculations required for convergence using best feature set $F_2$, worst feature set $F_6$, BMA$_1$ and BMA$_2$.}
\end{figure*}  

Similar to maximization of bulk modulus, the optimization for the minimization problem was carried out for feature sets $F_1,\ldots,F_6$, and then by using $BMA_1$ and $BMA_2$. The overall trend in the results was also similar: $F_2$ is found to be the best performing model on average, converging fastest to the minimum shear modulus. On the other hand, $F_6$ is uniformly the worst performing feature set on average, converging the slowest. The minimum shear modulus found in the experiment design iterations based on the best model ($F_2$), worst model ($F_6$), BMA$_1$, and BMA$_2$  averaged over all initial data instances are shown in Fig. \ref{fig:N_10_G_min_BMA} for $N=$10. The dotted line in the figure indicates the minimum shear modulus = 10.38 GPa that can be found in the MDS. The performance of both first-order and second-order BMA in identifying the minimum shear modulus lies close to that of the best single model ($F_2$).  Fig. \ref{fig:N_10_G_min_BMA_violinplot} shows the swarm plots corresponding to the results in Fig. \ref{fig:N_10_G_min_BMA}. It is seen that in almost 100 \% of the cases the optimal solution (minimum shear modulus) can be found within the designated budget when feature set $F_2$ is used, while very few instances of convergence are noted for $F_6$. Using BMA$_1$ and BMA$_2$ yields very satisfactory results, as a large majority of the cases converge within budget. Here, we see the advantage of using the BMA approach. Without having actually gone through the experiment design loop, one could not know \emph{a priori}, that using $F_6$ will result in not arriving at the desired material within a reasonable budget with a very high probability. This shows that if one were to just select a feature set even using domain knowledge, one may or may not select a good model. However, if one were to use the BMA approach, either BMA$_1$ or BMA$_2$, the probability of successfully arriving at the material with desired properties, is very high, since the BMA approach auto-selects the best feature set. 
Results for $N=$2,5,15,20 as well as the plots for BMA coefficients may be found in the Supplementary material\cite{supplementary}.

\subsection{Multi-objective optimization}

\subsubsection{Maximize bulk modulus and  Minimize shear modulus}

\begin{figure}
\centering
\includegraphics[width=0.975\columnwidth]{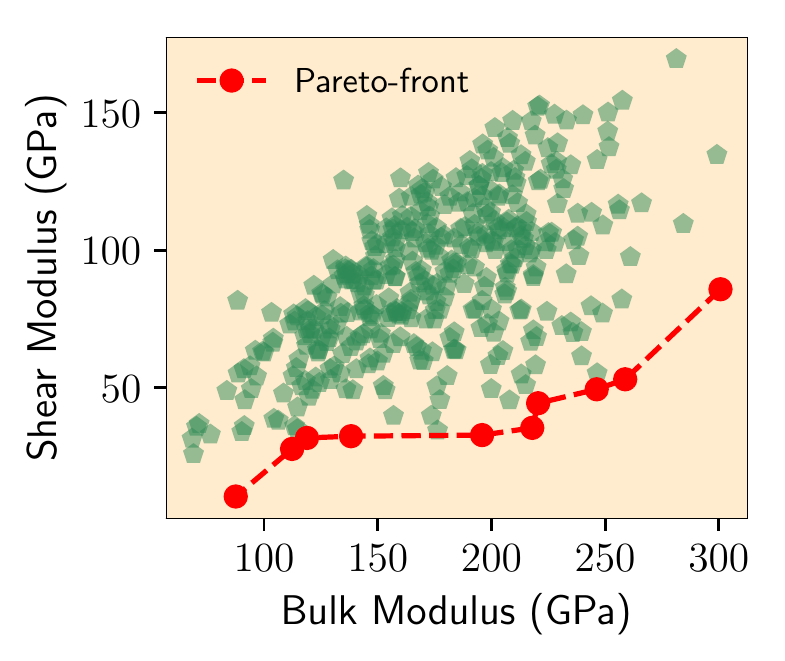}
 \caption{The Pareto optimal points in the materials property space are marked in red corresponding to the criterion of maximizing bulk modulus and minimizing shear modulus simultaneously. The Pareto set for the MDS consist of 10 points as indicated in red.}  
 \label{fig:pareto_front}
\end{figure}

We now consider multi-objective experiment design to optimize two objectives at the same time: maximizing bulk modulus and minimizing shear modulus. One should note that in our analysis we have already calculated the responses of bulk and shear modulus as materials properties for all the feasible points in the MDS to have the ground truth to compare different models for experiment design. Generally in practice, no knowledge of the responses exists unless one performs all the possible experiments exhaustively. Consequently, none of this information is used in our experiment design procedures. Fig. \ref{fig:pareto_front} illustrates all the data points in the objective space of materials properties (in green). It can be seen that in this case there does not exist a single optimal solution, and in fact there are 10 \emph{Pareto} optimal points comprising the \emph{Pareto front}\cite{ADEM:ADEM200300554} which is highlighted in red in the figure. Specifically, the Pareto front here is the 1-dimensional design curve
over which any improvement in one material property (i.e bulk modulus
K) is only achieved through a corresponding sacrifice of another
property (here, shear modulus G).

\begin{figure}
\centering
\includegraphics[width=0.975\columnwidth]{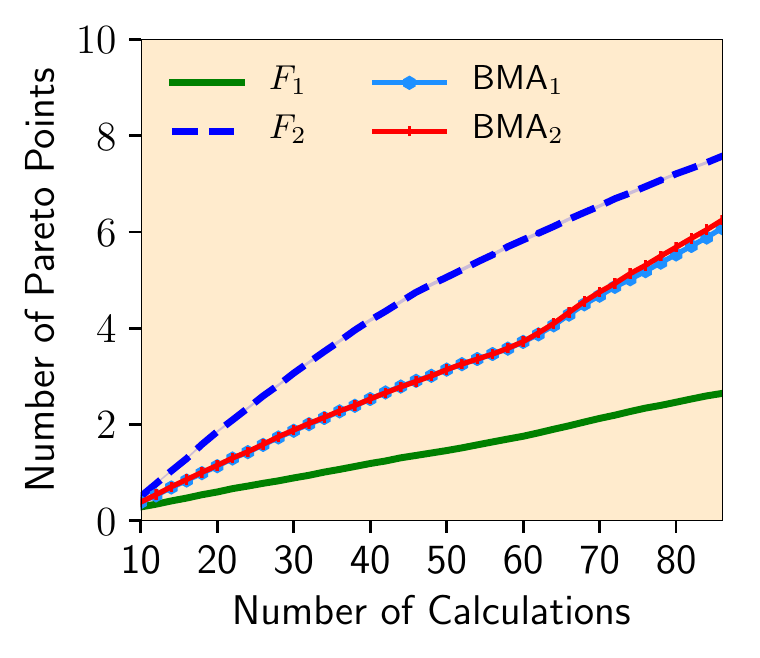}
 \caption{Average number of true Pareto optimal points found over all initial data set instances for single models, BMA$_1$, and BMA$_2$ for N=10}  
 \label{fig:N_all_pareto}
\end{figure}

Fig. \ref{fig:N_all_pareto} depicts the average performance of the best ($F_2$) and worst ($F_1$) models as well as the first- and second-order BMA in finding the true Pareto optimal points versus the number of calculations. Similar to single-objective problems, multi-objective experiment design based on $F_2$ consistently has the best performance; i.e. it identifies more true Pareto optimal points faster (with smaller budget). Both BMA approaches' performances are consistently in the range of the first best ($F_2$) single model's performances.  
Complete results for all cases of N, swarm plots and coefficient plots for the multi-objective scenario may be found in the Supplementary material \cite{supplementary}.

\section{Discussion}

\subsection{Comparison of first-order and second-order BMA}

From the results in the previous sections, we can see that for single-objective experiment design, the performance of the first-order BMA is slightly better than the second-order BMA. On the other hand, the model probabilities in the second-order BMA are more robust, and at any calculation number (sequential experiment iteration), the average posterior probability over all the initial data set instances of the best model in terms of experiment design performance is higher than the other models. The reason is that second-order Laplace approximation, unlike the first-order one, does not rely solely on the fitted values of the parameters of the GPR model to calculate the model probability. In fact, it approximates the model probability by integrating a local expansion of the marginal likelihood over a neighborhood of the fitted parameters values, which may dampen the fluctuations of the fitted values between different sequential experiment iterations. For the multi-objective case, the second-order BMA is slightly better than first-order BMA in terms of both experiment design performance and robustness of identifying the best model in terms of experiment design performance.

\begin{figure*}
        \parbox{.975\textwidth}{
            \begin{subfigure}{.475\linewidth}
                \includegraphics[width=\textwidth]{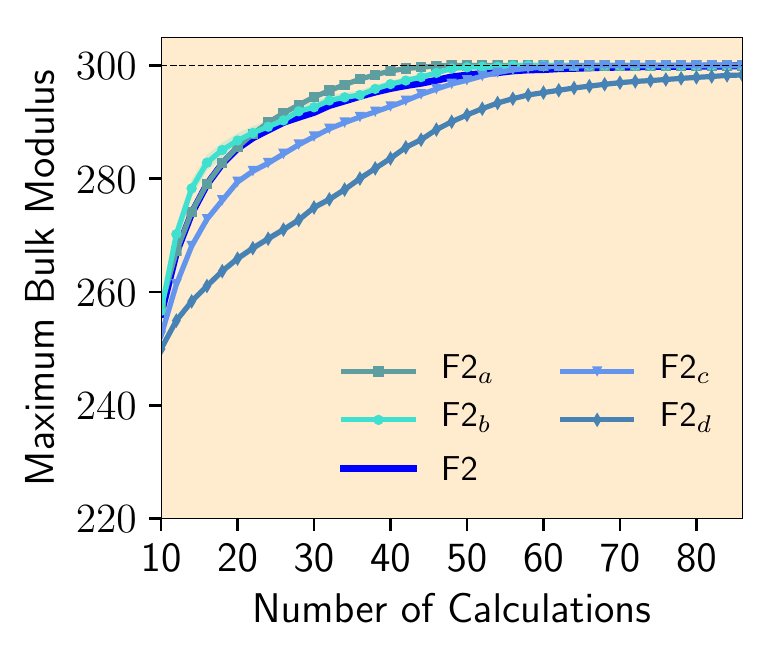}
                \caption{}
                \label{fig:max_bulk_modulus_combo_F2}
        \end{subfigure}
        \begin{subfigure}{.475\linewidth}
                \includegraphics[width=\textwidth]{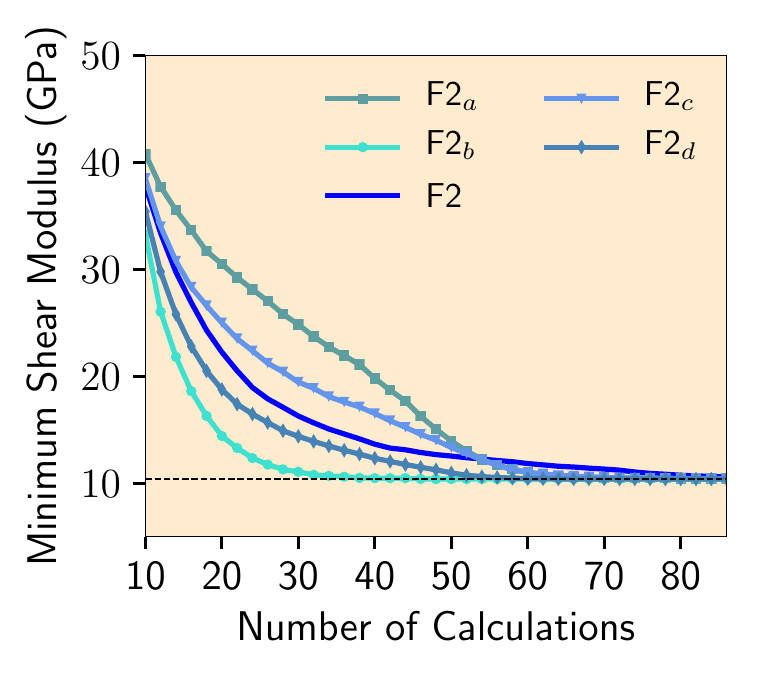}
                \caption{}
                \label{fig:min_shear_modulus_combo_F2}
        \end{subfigure}
        }
        \caption{Average maximum discovered a) bulk modulus and b) shear modulus for $F_2$ and lower-dimensional feature sets ($F_{2a}$,$F_{2b}$,$F_{2c}$,$F_{2d}$) derived from $F_2$}
         
\end{figure*}

\subsection{Remarks on feature sets}
The feature sets in our analysis are chosen \emph{a priori} based on domain knowledge. We do not claim that the considered feature sets are among the best possible feature sets for our experiment design problems. We are rather using these to showcase the applicability of the BOMU framework in real-world experiment design problems, where the best model or feature set is often not known, and only a set of possible models might exist based on domain knowledge. The power of BOMU is that it incorporates the uncertainty over the possible model space, instead of relying on a single model that is selected based on limited initial available data. For instance, we compared experiment design results based on the subsets of $F_2$ with one feature removed from $F_2$ (by taking three features at a time): feature set $F_{2a}$: $[m,Z,I_{dist}]$, feature set $F_{2b}$: $[m,Z,e_a]$, feature set: $F_{2c}$: $[m,I_{dist},e_a]$, and feature set $F_{2d}$: $[Z,I_{dist},e_a]$. 

Figs. \ref{fig:max_bulk_modulus_combo_F2} and~\ref{fig:min_shear_modulus_combo_F2} show the corresponding results for maximizing bulk modulus and minimizing shear modulus problems, respectively. From both figures, there are some subsets that can perform better in terms of average optimal objective values discovered over all instances of initial data sets for a fixed initial data set size.
Another observation from Figs. \ref{fig:max_bulk_modulus_combo_F2} and \ref{fig:min_shear_modulus_combo_F2} is that adding non-informative features to a model (feature set) can degrade the experiment design performance, as there are the single models based on some subsets of cardinality three derived from $F_2$  that can find the optimal compound in the MDS faster than the experiment design based on $F_2$. One reason is that by adding non-informative features, more dimensions are introduced in the feature space while the information on these dimensions may be irrelevant to their outputs---it does not help better predict the outputs. This has more effect especially when using kernels with a single length-scale parameter, which is the most common practice in the materials literature. This is explicitly indicated in Section \ref{sec:non_informative_features}, as the the BOMU approach excels when there are non-informative features, in that it auto-rejects feature sets F$_6$,..,F$_{10}$, while converging to the target experiment as fast as the best standalone model F$_2$.
Further discussion is included in Sec. \ref{App_A}

\section{Conclusions}
The Bayesian optimization approach was successfully combined with  Bayesian model averaging(BMA) for autonomous and adaptive learning to design a Bayesian experiment design framework under model uncertainty (BOMU) for materials discovery in single- and multi-objective material property space using a test set of MAX phases. It was demonstrated that, while prior knowledge about the fundamental features linking the material to the desired material property is certainly essential to build the Materials Design Space (MDS), the BMA approach may be used to auto-select the \emph{best} features/feature sets in the MDS, thereby eliminating the requirement of knowing the \emph{best} feature set \emph{a priori}. As evident from the extensive results included in the Supplementary material\cite{supplementary}, the BOMU framework is not significantly dependent on the size of the initial data, which enables its use in materials discovery problems where initial data is scant. At the very least, this framework provides a very efficient means of building the initial data set as well, since it may be used to guide experiments or calculations by focusing on gathering data in those sections of the MDS which will result in the most efficient path to achieving the optimal material.

\section{Acknowledgments}
\label{sec:ack}
The authors acknowledge the support of NSF through the project \emph{DMREF: Accelerating the Development of Phase-Transforming Heterogeneous Materials: Application to High Temperature Shape Memory Alloys}, NSF-CMMI-1534534. RA and ED also acknowledge the support of NSF through Grant No. NSF-DGE-1545403. TD acknowledges support of NSF-CMMI-1729335. XQ acknowledges the support of NSF through the project \emph{CAREER: Knowledge-driven Analytics, Model Uncertainty, and Experiment Design}, NSF-CCF-1553281 and NSF-CISE-1835690(with RA). AT and RA also acknowledge support by the Air Force Office of Scientific Research under AFOSR-FA9550-78816-1-0180 (Program Manager: Dr. Ali Sayir).

AT and SB contributed equally to this work.

\section{Appendix}
\subsection{Implementation Remark}
\label{App_A}
The estimation of the (hyper) parameters, including the length-scale parameter, of the GPR model are found by maximizing the marginal likelihood of the data, i.e. ML-II estimation instead of the fully Bayesian treatment. Marginal likelihood might have multiple optima that correspond to different interpretations of the data. When GPR models are trained based on ML-II estimation, depending on the MDS and selected kernel functions, there is a possibility of overfitting the training data, especially when only a small number of measured data points are available (small-sample training data as initial data points). One thing to note is that experiment design based on GPR models that overfit the training data and assign very low correlation to nearby points in their prediction can yield very poor experiment design performance. One reason being that in this case measuring any point in the MDS will not give much information regarding other points of the MDS, because of the overfitting of the underlying learned surrogate GPR model. Since in our experiments the feature sets were chosen \emph{a priori}, without any knowledge of their suitability for the underlying true model that generates data, in our implementation we have restricted the possible range for the length-scale parameter of the GPR kernel to prevent the models from overfitting the limited number of available data.
\subsection{Connections and Differences with Generalized MOCU}
\label{sec:addedappendix} We would like to close with some remarks
concerning the manner in which the experiment design developed in this paper
relates to the general theory. In the following we first provide a brief
summary of the generalized MOCU introduced in \cite{shahin2018}. Assuming a
probability space $\mathcal{M}$ (uncertainty class) with probability measure 
$P$, an action space $\mathcal{X}$, and an objective function $f:\mathcal{M}%
\times \mathcal{X}\rightarrow (-\infty ,\infty )$, our goal is to find an
action $\mathbf{x}\in \mathcal{X}$ that minimizes the unknown true objective
function $f(\mathbf{x};M_{t})$ over $\mathcal{X}$, where $M_{t}\in \mathcal{M%
}$. A \emph{robust action} is an element $\mathbf{x}^{R}\in \mathcal{X}$
that minimizes the average of the objective function across all
possibilities in the uncertainty class relative to a probability
distribution governing the corresponding space. This probability at each
time step is the posterior distribution given the observed data points
available up to that time step ($\mathcal{D}_{n}$). Mathematically,

\begin{equation}
\mathbf{x}_{n}^{R}=\underset{\mathbf{x}\in \mathcal{X}}{\argminA}E_{M}[f(%
\mathbf{x};M)|\mathcal{D}_{n}].
\end{equation}%

The mean objective cost of uncertainty (MOCU) is the average gain in the
attained objective between the robust action and the actual optimal actions
across the possibilities:

\begin{equation}
\text{MOCU}_{n}^{\mathcal{X}}(\mathcal{M})=E_{M}[f(\mathbf{x}_{n}^{R};M)-f(%
\mathbf{x}_{M}^{\ast };M)|\mathcal{D}_{n}],
\end{equation}
where $\mathbf{x}_{M}^{\ast }$ denotes the optimal action for a given $M$.
Note that if we actually knew the true (correct) model, then we would simply
use the optimal action for that model and MOCU would be 0. Denoting the set
of possible experiments by $\Xi $, the best experiment $\xi _{n}^{\ast }$ at
each time step (in one step look ahead scenario) is the one that maximally
reduces the expected MOCU following the experiment, i.e. 
\begin{equation}
\begin{split}
\xi _{n}^{\ast }=\underset{\xi \in \Xi }{\argminA}E_{\xi }[& E_{M}[f(\mathbf{%
x}_{n+1}^{R};M)|\xi ,\mathcal{D}_{n}]] \\
& -E_{M}[f(\mathbf{x}_{n}^{R};M)|\mathcal{D}_{n}].
\end{split}
\label{eq:remaining_MOCU_data}
\end{equation}%
In most cases in the context of materials discovery, each experiment is
applying an action and observing its cost (or a noisy version of it). Thus,
the experiment space is equivalent to the action space.

It is beneficial to recognize that MOCU can be viewed as the minimum
expected value of a Bayesian loss function, where the Bayesian loss function
maps an action to its differential objective value (for using the given
action instead of an optimal action), and its minimum expectation is
attained by an optimal robust action that minimizes the average differential
objective value. In decision theory, this differential objective value has
been referred to as the \emph{regret}.

In Section \ref{sec:addedsubsection}, we mentioned three possibilities
regarding the objective function. In the first case, we have a parametric
model where the parameters come from an underlying physical system. An
example in medicine is where they characterize a gene regulatory network,
the objective function is the likelihood of the cell being in a cancerous
state, and the action is to administer a drug \cite{Dehghannasirimocu}.
Another example is in imaging where the parameters characterize the image
structure, the objective function is an error measure between two images,
and the action is to compress the image in order to reduce the number of
bits while at the same time maintaining visual fidelity \cite{roozbeh2018}. In this case
the action space and experiment space are usually distinct sets.

Another possibility is that the features are known and the parameters come
from a surrogate model used in place of the actual physical model, but
believed to be appropriately related to the physical model. In the materials
example \cite{RoozbehMaterial} noted in Section \ref{sec:addedsubsection}, the surrogate model is
based on the time-dependent Ginzburg-Landau (TDGL) theory and simulates the free energy given dopant parameters,
the objective function is the energy dissipation, and the action is to find
an optimal dopant and concentration. To see how the
approach in \cite{RoozbehMaterial} fits the above general theory the reader
can refer to \cite{shahin2018}.

A third possibility is that we do not know the physical model and we lack
sufficient knowledge to posit a surrogate model with known features/form
relating to our objective. This case arises in many scenarios where the
objective function is a black box function. Nevertheless, we can adopt a
model, albeit, one with known predictive properties. This model can be a
kernel-based model like a GP. Moreover, this model can consist of a set of
possible parametric families, or a kernel-based model with different
possible feature sets, or even kernel-based models with different choices
for the kernel function. In this paper, we have addressed this case when we
do not \emph{a priori} have any knowledge about which feature set or model
family would be the best, and reliable model selection cannot be performed
before starting the experiment design loop. Considering the average
prediction from models based on different feature sets or model families
weighted by their posterior probability of being the correct model, namely
BMA, is one possible approach. In this paper we perform BMA based on
possible feature sets that come from domain knowledge.

It is worth mentioning that, in theory, the generalized MOCU can be applied
to all these scenarios with a single objective; however, there might be
computational issues, especially in the third type of model. For example,
when the experiments consist of running expensive simulation models, the
computations of MOCU-based experiment design might be extremely heavy, so
much so that the experiment design would be more computationally expensive
and/or time consuming than the original simulation model.

A last question needs to be addressed. As noted previously, it is known that
under certain conditions, MOCU-based experiment design is equivalent to EGO 
\cite{shahin2018}. Could we have used MOCU here, and/or can the procedure
proposed in this paper be related to MOCU? In our case, at each time step,
after training the GPs based on the current and previous observations
(finding the GP hyperparameters that maximize the marginal likelihood of the
observed data), each GP provides a Gaussian distribution over the objective
values of the actions. Averaging several GPs based on their posterior model
probabilities is like mixing weighted Gaussian distributions over the
objective value of each action. Based on the sum of weighted Gaussian
distributions, the EI or EHVI is calculated for all possible remaining
actions for single- or multi-objective scenarios, respectively, and the
maximizer is chosen as the next experiment. For the multi-objective case, we
can not employ MOCU. The reason is that the current formulations of MOCU do
not contain definitions suitable to multi-objective problems, e.g. no notion
of robust action exists in the presence of Pareto optimal solutions. For the
single-objective case, assuming the mixture of Gaussian distributions for
the objective value of each action given at each time step, and confining
the selection of the optimal action in the MOCU framework at each time step
to the set of actions whose objective values have been previously observed,
the maximizer of EI is equivalent to the solution of %
\eqref{eq:remaining_MOCU_data}. But in practice we have another layer of
uncertainty introduced by the model fitting step. If we want to take this
uncertainty into account when calculating the expected utility (acquisition
value) at each time step, the procedure taken in this paper by employing EI
is not equivalent to applying MOCU. To make it so we would have to assume a
prior distribution over the hyperparameters of the GPs and when calculating
the expected utility (acquisition value) of each potential next experiment
at each time step, we would have to consider the corresponding possible
updated distributions of the hyperparameters and consequent model
probabilities posterior to carrying out the experiment and the possible
objective value observation in the next time step. But this would be too
computationally costly. 

\end{document}